\documentclass[aps,twocolumn,prc]{revtex4-2}

\usepackage{graphicx}
\usepackage{tabularx}
\usepackage{dcolumn}
\usepackage{nicefrac}
\usepackage{braket}

\newcommand{\Bhf}{\ensuremath{B_{\mathrm{hf}}}}

\begin{document}

\title{Hyperfine fields at $^{\mathbf{66}}$Ga, $^{\mathbf{67,69}}$Ge implanted into iron and gadolinium hosts at 6~K, and applications to $\mathbf{g}$-factor measurements}

\author{T.~J.~Gray}
\author{A.~E.~Stuchbery}
\author{B.~J.~Coombes}
\author{J.~T.~H.~Dowie}
\author{M.~S.~M.~Gerathy}
\author{T.~Kib\'edi}
\author{G.~J.~Lane}
\author{B.~P.~McCormick}
\author{A.~J.~Mitchell}
\author{M.~W.~Reed}
\affiliation{Department of Nuclear Physics, Research School of Physics, Australian National University, Canberra ACT 2601, Australia}

\date{\today}

\begin{abstract}
Isomers in $^{66}$Ga, $^{67}$Ge, and $^{69}$Ge were recoil-implanted into ferromagnetic hosts of iron and gadolinium at $\approx 6$~K, and the hyperfine magnetic fields were determined by Time Differential Perturbed Angular Distribution (TDPAD) measurements. The hyperfine field strengths at $\approx 6$~K are compared to the results of previous higher-temperature measurements and the amplitudes of the $R(t)$ functions are compared to empirical expectations.  The results show that gadolinium can be a suitable host for high-precision in-beam $g$-factor measurements. The results of new $g$-factor measurements for isomers in $^{66}$Ga and $^{67}$Ge are $g(^{66}$Ga$,7^{-}) = +0.126(4)$, supporting a $[\pi_{f5/2} \otimes \nu g_{9/2}]_{7^-}$ configuration assignment, and $g(^{67}$Ge$,\nicefrac{9}{2}^{+})=-0.1932(22)$, derived from a new measurement of the ratio $g(^{67}\mathrm{Ge})/g(^{69}\mathrm{Ge}) = 0.869(9)$. These values are in agreement with previous results. The $R(t)$ amplitudes indicate that the nuclear alignment produced in the isomeric states was significantly lower than the empirically expected $\sigma/I \approx 0.35$. %
\end{abstract}

\maketitle

\section{Introduction}
\paragraph*{}
Nuclear magnetic moments can provide valuable insight into the single-particle nature of excited nuclear states; see, for example, Refs.~\cite{Brown2004,Mertzimekis2003,Kusakari1984,Ingwersen1975,Mucher2009}.  Many techniques use the spin-precession of an excited nuclear state to measure its magnetic moment. In order to probe shorter-lived states ($<100$~ns) and/or small $g$-factor values, intense magnetic fields ($\geq 20$~T) must be used~\cite{Morinaga1976,Raghavan1985,Stuchbery2019}.  In view of this requirement, the static hyperfine field ($\Bhf$) that a nucleus experiences as a dilute impurity inside a ferromagnetic host is an important tool~\cite{Morinaga1976,Raghavan1985}.  By recoil-implanting the nucleus of interest into a ferromagnet, the static hyperfine field can be used to measure nuclear moments in-beam, in principle enabling the study of a wide variety of nuclear states by both time-integral and time-differential methods~\cite{Fahlander1979_2,Rafailovich1983,Stuchbery2018,Hensler1971}.

\paragraph*{}
Iron is often used as a ferromagnetic host due to its face-centered cubic structure (no electric quadrupole interactions)~\cite{Mohn2000,Raghavan1985}, high internal field strengths, and high Curie temperature of $T_c = 1043$~K.  However, in the context of in-beam studies, ferromagnetic materials with high $Z$ such as gadolinium ($T_c = 293$~K) are advantageous, as higher beam energies can be used without exceeding the Coulomb barrier, and inducing high levels of unwanted background radiation.%

\paragraph*{}
In a recent publication we measured the hyperfine field of $^{107}$Cd recoil-implanted into gadolinium using LaBr$_3$ detectors and the Time Differential Perturbed Angular Distribution (TDPAD) method~\cite{Gray2017}. The objective in that work was to re-examine the $g$ factor of the yrast $10^+$ state in $^{110}$Cd, which had been measured previously by the integral perturbed angular distribution (IPAD) method~\cite{Regan1995} following implantation into a gadolinium host, and was expected to have a relatively pure $\nu (h_{\nicefrac{11}{2}})^2$ configuration.  The inferred $g$ factor in Ref.~\cite{Regan1995} was considerably reduced in magnitude from the $g \approx -0.2$ suggested by empirical $g$ factors observed for $\nu h_{\nicefrac{11}{2}}$ isomers in the odd-$A$ Cd isotopes near $A=110$~\cite{ENSDF}.  Ref.~\cite{Gray2017} measured the time-dependent perturbed angular distribution of the $640$-keV transition depopulating the $846$-keV, $I^{\pi}=\nicefrac{11}{2}^-$ isomer in $^{107}$Cd following the $^{98}$Mo$(^{12}$C$,3\mathrm{n})$ reaction.  This perturbed angular distribution showed that a large fraction of the $^ {107}$Cd nuclei implanted into the gadolinium host experienced a very low (effectively zero) hyperfine field, although most of the remainder experienced a field of $\approx34$~T near the $|\Bhf|=34.0(7)$~T expected for Cd in gadolinium from offline measurements~\cite{Forker1973}.  This observation allowed a reevaluation of the IPAD data and resolved the discrepancy between the expected and measured $g(10^+)$ values in $^{110}$Cd. Ref.~\cite{Gray2017} concluded that caution is required if time-integral \mbox{$g$-factor} measurements are performed since only the average precession angle from all implantation sites is observed.  Time-dependent measurements do not suffer from this problem, and are sensitive to the distribution of fields after implantation, making them a more suitable tool in this context.

\paragraph*{}
However, our measurements on $^{107}$Cd implanted into gadolinium~\cite{Gray2017} also raised concerns about the utility of gadolinium as a ferromagnetic host for time-differential $g$-factor measurements. Specifically: (i) the static hyperfine field strength was not single valued and it was instead distributed over a range of strengths, and (ii) of more concern, the distribution of fields changed with time.  Accumulating radiation damage caused by the high beam intensities used in that work was suggested as the cause of this variation~\cite{Gray2017}.

\paragraph*{}
In view of this previous experience, the present work sought to investigate the utility of gadolinium hosts for TDPAD $g$-factor measurements under conditions where the beam intensity is not excessive, the fraction of implanted ions on full-field sites can be evaluated, and the reproducibility of the measurements can be tested.
Our aim was to understand the behaviour of gadolinium as a host material, and to use LaBr$_3$ detectors in-beam for TDPAD moment measurements following recoil implantation. The ultimate objective is to gain access to $g$-factor measurements on a class of isomers that have not been accessible to experiment previously due to their short lifetimes and/or short precession periods ($\approx 10$~ns). The present measurements do not strive for this short-lifetime limit, but focus more generally on the behavior of hyperfine fields following implantation into gadolinium hosts.
To this end, we measured time dependent perturbed angular distributions following the implantation of $^{67}$Ge into gadolinium, as well as for $^{67,69}$Ge and $^{66}$Ga into iron hosts. These isomers were chosen because they are strongly populated in convenient reactions, are well characterized, and because they have been used for relevant similar measurements performed independently in other laboratories. All measurements were performed at $\approx 6$~K.  The measurements with iron hosts serve to benchmark the case where all implantations have effectively $100\%$ substitutional (full-field) sites, and thus provide a baseline by which to assess the field-free fraction in the case of the gadolinium host. The evidence for near $100\%$ substitution of Ga and Ge ions on full-field sites from the literature~\cite{Raghavan1985,Lee1991} and the present work is discussed below (Sec.~\ref{sec:6769_ge_fe}).

\paragraph*{}
The present work is compared with previous measurements~\cite{Raghavan1978,Raghavan1979,Raghavan1985,Lee1991,Filevich1978}, most at higher temperatures ($77$~K -- $300$~K), but all well below the Curie temperatures of the hosts ($T_c = 1043$~K for iron and $T_c = 293$~K for gadolinium).  These comparisons allow an evaluation of the reproducibility of TDPAD measurements with gadolinium hosts in different laboratories, given that the temperature dependence of the hyperfine fields can be shown to be small in the temperature range considered~\cite{Bozorth1993,Bernas1973,Fahlander1979}.

\paragraph*{}
As a by-product of these experiments we have remeasured the $g$ factor of the $7^{-}$ isomer in $^{66}$Ga, and the $g$-factor ratio $g(\nicefrac{9}{2}^{+}$ $^{67}$Ge$)/g(\nicefrac{9}{2}^{+}$ $^{69}$Ge$)$.  Furthermore, new information on the temperature dependence of the hyperfine fields for Ge in iron and gadolinium has been obtained.

\section{Experimental Method}
\paragraph*{}
Four experimental runs took place at the Heavy Ion Acceleartor Facility (HIAF) at the Australian National University (ANU).  The 14UD Pelletron accelerator provided pulsed 48-MeV $^{12}$C and 60-MeV $^{16}$O beams to populate isomers in $^{66}$Ga, $^{67}$Ge, and $^{69}$Ge.  Reaction and target details are listed in Table~\ref{tab:targets}. Beams were pulsed in bunches of $\approx 2$~ns full width at half maximum (FWHM).  The separations between beam pulses are given in the table. Beam intensities were near 1~pnA. Properties of the isomers investigated are summarized in Table~\ref{tab:isomers}.

\begin{table*}[]
  \centering
  \caption{Reaction and target details.}
  \label{tab:targets}
  \begin{ruledtabular}
    \begin{tabular}{llllll}
  Run & Reaction                                       & Energy & First Layer              & Second Layer               & Pulse Separation \\
      &                                                & (MeV)  & (mg/cm$^2$)              & (mg/cm$^2$)                & (ns)             \\ \hline
  1   & $^{56}$Fe($^{12}$C, pn)$^{66}$Ga               & 48     & 3.44 $^{\mathrm{nat}}$Fe & none                       & 972              \\
  2   & \mbox{$^{54,56}$Fe($^{16}$O, 2pn)$^{67,69}$Ge} & 60     & 5.09 $^{\mathrm{nat}}$Fe & none                       & 9872             \\
  3   & $^{54}$Fe($^{16}$O, 2pn)$^{67}$Ge              & 60     & 1.16  $^{54}$Fe          & $5.09$ $^{\mathrm{nat}}$Fe & 972              \\
  4   & $^{54}$Fe($^{16}$O, 2pn)$^{67}$Ge              & 60     & 1.16 $^{54}$Fe           & 3.2 $^{\mathrm{nat}}$Gd    & 972              \\
    \end{tabular}
    \end{ruledtabular}
\end{table*}

\begin{table*}[]
  \centering
  \caption{Properties of nuclear isomers investigated in the present work.}
  \label{tab:isomers}
  \begin{tabularx}{\textwidth}{XXXXXXX} \hline \hline
    Nuclide   & $E_x$ (keV) & $I^\pi$             & $\tau$ (ns) & $E_\gamma$ (keV) & $g$           & Run/[Ref.]                \\\hline
    $^{66}$Ga & $1464$      & $7^-$               & $83$        & $601$            & $+0.126(4)$   & 1                       \\
    $^{69}$Ge & $398$       & $\nicefrac{9}{2}^+$ & $4050$      & $398$            & $-0.2224(7)$  & \cite{Christiansen1970} \\
    $^{67}$Ge & $752$       & $\nicefrac{9}{2}^+$ & $160$       & $734$            & $-0.1932(22)$ & 2                       \\ \hline\hline
\end{tabularx}
\end{table*}

\paragraph*{}
All host foils in the present and previous work in our laboratory were prepared by cold-rolling and then annealing under vacuum at \mbox{$\approx 850$~$^\circ$C} for 20 minutes. Gadolinium foils were rolled from $99\%$ pure $25$~$\mu$m foils obtained from Goodfellow. By sandwiching the gadolinium foils between clean tantalum sheets, which act as a getter when heated in vacuum, the risk of surface contamination during annealing is effectively eliminated.  Iron foils were annealed in a similar manner. The experiment was conducted using the ANU Hyperfine Spectrometer~\cite{Stuchbery2019}. A \mbox{$\approx 0.1$-T} field provided by an electromagnet polarized the ferromagnetic host foil in the vertical direction.  The field direction was reversed every $\approx 15$ minutes to minimize systematic uncertainties. The targets were cooled by a cryocooler maintained at $5.7(1)$~K throughout the runs. We report here the temperature measured on the target frame.  As discussed in Ref.~\cite{Stuchbery2019}, the temperature in the beam spot could be somewhat higher. Calculations assuming $1$~pnA of beam current indicate that shifts in temperature of as much as 20~K may be possible. However, because the Brillouin function is flat near $0$~K (see Fig.~\ref{fig:brillouin} below), an increase to even 30~K or $0.1T_c$ for gadolinium does not materially affect the following discussion~\cite{Bozorth1993,Bernas1973,Fahlander1979}.

\paragraph*{}
Four $\gamma$-ray detectors were placed in the horizontal plane perpendicular to the magnetic field direction. In runs 1, 2, and 4, two HPGe detectors were located at $\theta_\gamma = \pm135^\circ$ relative to the beam axis, and two LaBr$_3$ scintillators were located at $\theta_\gamma = \pm45^\circ$. In run 3, four LaBr$_3$ detectors were located at $\theta_\gamma = \pm45^\circ$ and $\theta_\gamma = \pm135^\circ$. The HPGe detectors had timing resolution with FWHM of $<10$~ns at $\approx 500$~keV.  The LaBr$_3$ detectors had far superior timing resolution, with FWHM negligible in comparison to the 2~ns timing spread of the beam pulse. Since the oscillations in the present work all have periods $>20$~ns, in all cases the timing resolution is much shorter than the precession periods being observed. Ortec 567 Time to Amplitude Converters (TACs) recorded times relative to the beam pulse. TAC ranges were $1$~$\mu$s for run 1, 3, and 4, where isomer lifetimes $\approx 100$~ns were studied.  When $^{69}$Ge was present with significant intensity (run 2), all TACs were set on the \mbox{$10$-$\mu$s} range to allow the isomeric state with $\tau = 4.05$~$\mu$s to be measured over two mean lives.

\section{Analysis procedure}
%%%%%%%%%%%%%%%%%%%%%%%%%
% energy spectra for Ga %
%%%%%%%%%%%%%%%%%%%%%%%%%
\begin{figure}
  \centering
  \includegraphics[width=0.5\textwidth]{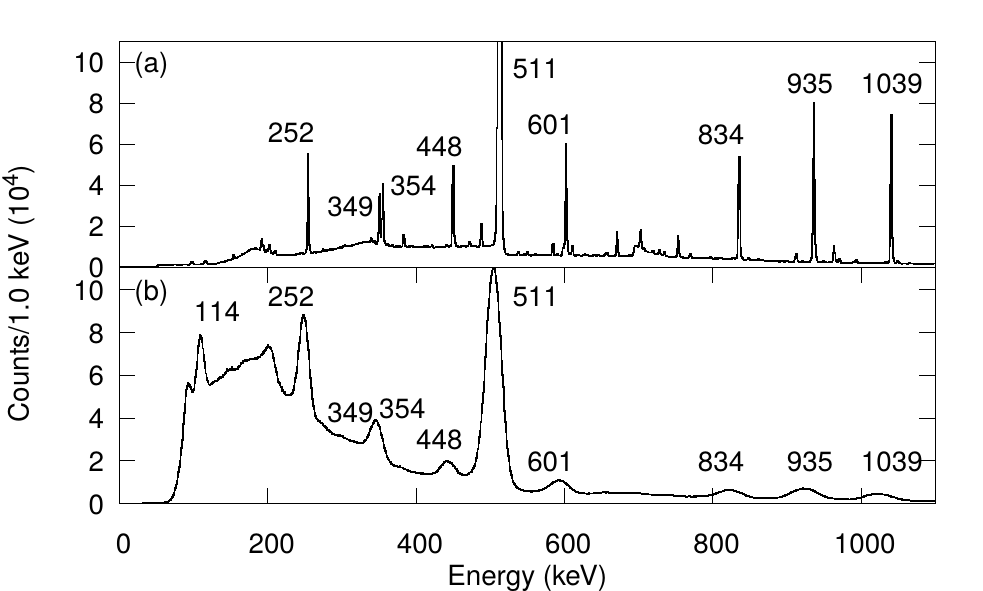}
  \caption{Out-of-beam energy spectra following $^{12}$C-induced reactions on $^{\mathrm{nat}}$Fe to populate $^{66}$Ga, as observed in (a) HPGe and (b) LaBr$_3$ detectors. The 834-keV and 1039-keV transitions are $^{66}$Zn activity after the decay of $^{66}$Ga. All other strong transitions follow from the depopulation of the $I^{\pi} = 7^{-}$ isomeric state in $^{66}$Ga.}
  \label{fig:ga_espectra}
\end{figure}

%%%%%%%%%%%%%%%%%%%%%%%%%
% energy spectra for Ge %
%%%%%%%%%%%%%%%%%%%%%%%%%
\begin{figure}
  \centering
  \includegraphics[width=0.5\textwidth]{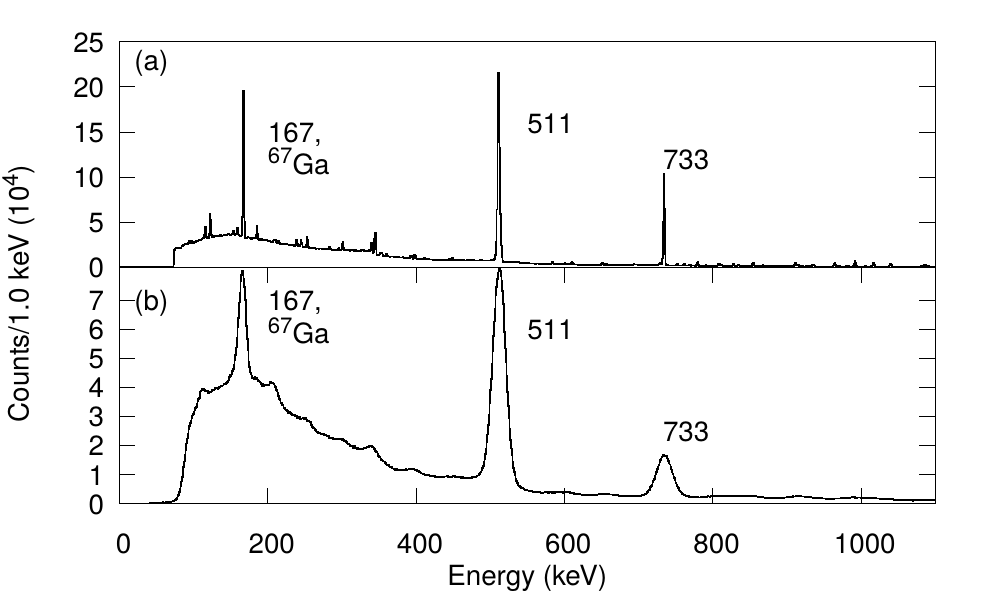}
  \caption{Out-of-beam energy spectra following $^{16}$O-induced reactions on $^{54}$Fe to populate the $\nicefrac{9}{2}^+$ isomer in $^{67}$Ge. These spectra are from run 3, where a gadolinium foil is the ferromagnetic host; (a) shows a spectrum from HPGe detectors, and (b) from LaBr$_3$ detectors.}
  \label{fig:ge_espectra}
\end{figure}

\begin{figure}
  \centering
  \includegraphics[width=0.5\textwidth]{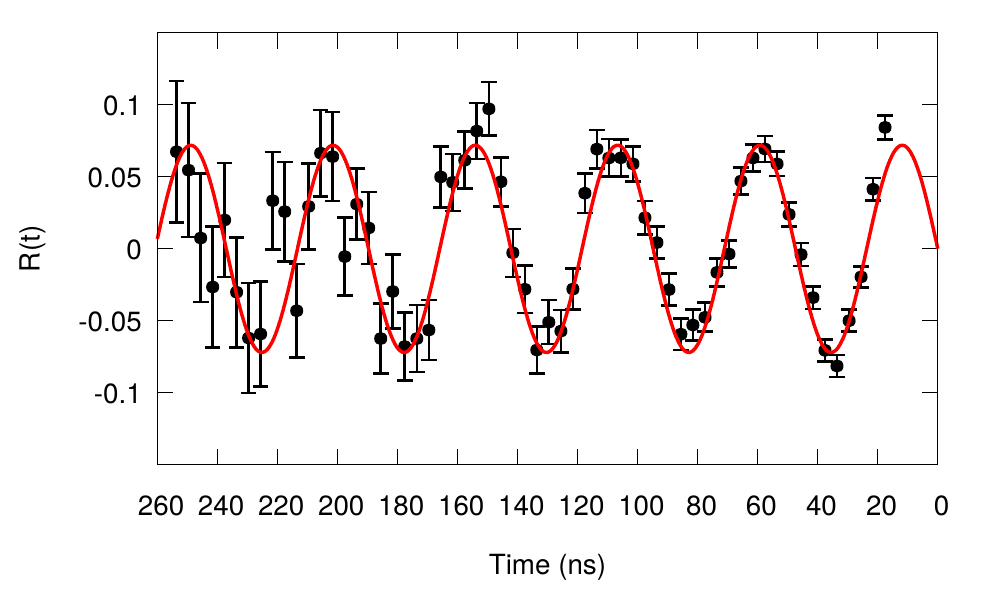}
  \caption{$R(t)$ function for $^{66}$Ga implanted into iron (run 1). $\chi^2/\nu = 1.37$.}
  \label{fig:gafe}
\end{figure}

\paragraph*{}
Out-of-beam $\gamma$-ray energy spectra are shown in Figs.~\ref{fig:ga_espectra} and \ref{fig:ge_espectra} following the $^{56}$Fe($^{12}$C,pn) reaction in run 1 and $^{54}$Fe($^{16}$O,2pn) in run 3, respectively. These spectra show transitions depopulating the isomeric states detected in both LaBr$_3$ and HPGe detectors.  All major lines have been identified.

\paragraph*{}
Beam-$\gamma$ time spectra are formed by gating on the relevant depopulating transitions, $E_{\gamma}$, as listed in Table~\ref{tab:isomers}.  The time spectra formed by gating on adjacent peak-free regions of the energy spectra are then subtracted, to form background-subtracted time spectra. These show oscillations in time due to the precession of the excited state, in addition to the exponential decay of the isomer. Two such histograms were formed for each case studied, corresponding to different field-direction/detector combinations that have oscillations out of phase with each other. Explicitly: one histogram, $N_1$, contains counts from detectors at $\theta_\gamma = +45^\circ$ and $\theta_\gamma = -135^\circ$ when the field is up, and counts from detectors at $\theta_\gamma = -45^\circ$ and $\theta_\gamma = +135^\circ$ when the field is down, whilst the other histogram $N_2$ contains the corresponding opposite combinations. As shown in Refs.~\cite{Georgiev2006,Recknagel1974}, these histograms can be used to construct an $R(t)$ function:
\begin{equation}
  R(t) = \frac{N_1(t) - N_2(t)}{N_1(t) + N_2(t)} \approx \frac{3a_{2}}{4+a_{2}}\sin(2\omega_L t),
\end{equation}
where $N_i(t)$ represents the counts in each histogram, $a_{2} = A_2B_2$, where $B_2$ is the statistical tensor specifying the spin-alignment of the isomeric state, and $A_2$ is the appropriate combination of $F_2$ coefficients associated with the spin change and multiplicity of the $\gamma$-ray decay~\cite{Stuchbery2003,Yamazaki1967}. In other words, the unperturbed angular correlation has the form \mbox{$W(\theta) = 1 + a_2P_2(\cos(\theta_{\gamma})) + a_4P_4(\cos(\theta_{\gamma}))$}, and it is assumed that $a_4 \approx 0$ as is usually the case following fusion-evaporation reactions~\cite{Recknagel1974}; see for example the $\gamma$-ray spectroscopy study of $^{69}$Ge by Zobel~\textit{et al.}~\cite{Zobel1979}. The Larmor precession frequency $\omega_L$ is:
\begin{equation}
  \omega_L = -g \frac{\mu_N}{\hbar} \Bhf,
\end{equation}
where $g$ is the nuclear $g$ factor, $\mu_N$ is the nuclear magneton, $\hbar$ is the reduced Plank constant, and $\Bhf$ the hyperfine field strength. Thus an experimental $R(t)$ function can be used to extract $\omega_L$, which is related to the quantities of interest $\Bhf$ and $g$.
This procedure was carried out to form $R(t)$ functions for each of the cases studied.%
\paragraph*{}
It is worth noting for the discussion below in Sec.~\ref{sec:discussion_amplitude} that the observed amplitude of the $R(t)$ function can be related to the $a_{2}$ coefficient, and hence the $B_2$ statistical tensor that specifies the spin-alignment of the isomeric state. However, if a significant proportion, $p$, of implantations are on low- or field-free sites, the amplitude of $R(t)$ will decrease correspondingly:
\begin{equation}
  R(t) \approx (1-p)\frac{3a_{2}}{4+a_{2}}\sin(2\omega_L t).
\end{equation}
Hence, the field-free fraction, $p$, can be estimated from the initial $R(t)$ amplitude, provided that the initial spin-alignment, $B_2$, is known. %
\paragraph*{}
The nuclear alignment following fusion-evaporation reactions is often specified by an oblate Gaussian distribution of $m$-substates~\cite{Yamazaki1967}. Typically, the standard deviation of this distribution is $\sigma \approx 0.35 I$, where $I$ is the angular momentum of the excited nuclear state~\cite{Stuchbery2002,Carpenter1990,Grau1974,Simms1974}. This model can serve to estimate $B_2$ and hence $a_{2}$.  Alternatively, the alignment $B_2$ can be estimated from measured angular distributions of transitions depopulating other (non-isomeric) states near the isomer of interest~\cite{Gray2017}. Direct measurements of the angular distributions from isomeric states are difficult to interpret because they can be attenuated by uncontrolled hyperfine interaction effects such as electric field gradients in the host.
\paragraph*{}%
If $\Bhf$ is not single valued, and instead distributed over a range of strengths, the amplitude of the $R(t)$ function will attenuate over time, at a rate proportional to the width of the $\Bhf$ distribution.  This rate of attenuation can allow the $\Bhf$ distribution width to be quantified in terms of an average field with a distribution about the average.  Gaussian distributions have been used in the literature~\cite{Raghavan1979}. However, alternative shaped distributions can equally well describe the data in some cases~\cite{Gray2017}.

\section{Results}
\subsection{$^{\mathbf{66}}$Ga in iron}
Figure \ref{fig:gafe} shows the $R(t)$ function obtained for $^{66}$Ga implanted into iron. This ratio function shows negligible time-dependent attenuation of its amplitude, indicating that $\Bhf$ has a single value, or a very narrow distribution of fields near the dominant central value.  The solid line in Fig.~\ref{fig:gafe} is the result of a fit which assumes a single $\Bhf$ value. The free parameters are the Larmor precession frequency, the $B_{2}$ coefficient, and the time-zero position. Using $\Bhf = 11.0(3)$~T from a NMR measurement by the spin-echo method at $4.2$~K~\cite{Kontani1965}, the observed $\omega_L$ can be used to deduce the $g$ factor of the isomeric state, to give $g = +0.126(4)$.  This value agrees well with previous measured values of $g = +0.127(3)$~\cite{Raghavan1985}, $g = +0.125(4)$~\cite{Filevich1978}, and $g = +0.123(30)$ (this value is from a $4$~T external field measurement from an unpublished thesis, quoted in~\cite{Filevich1978}).

\subsection{$^{\mathbf{67,69}}$Ge in iron}
\label{sec:6769_ge_fe}
\begin{figure}
  \centering
  \includegraphics[width=0.5\textwidth]{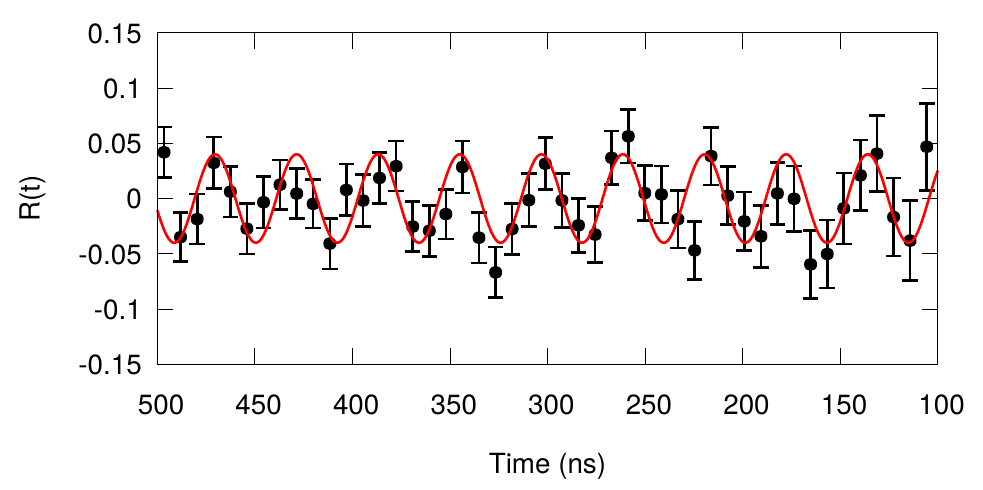}
  \caption{A section of the $R(t)$ function for $^{69}$Ge in iron, using the $^{\mathrm{nat}}$Fe target (run 2). $\chi^2/\nu = 1.12$.}
  \label{fig:ge69_ratio}
\end{figure}

\begin{figure}
  \centering
  \includegraphics[width=0.5\textwidth]{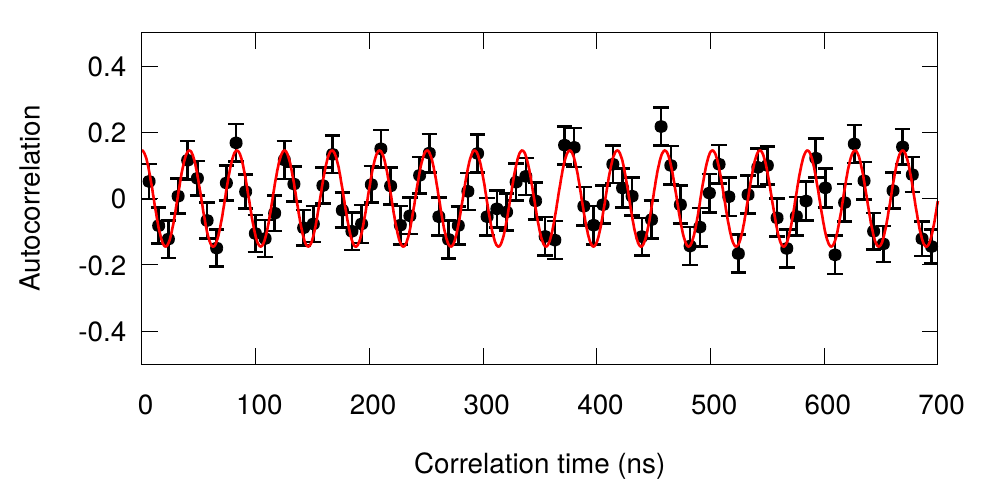}
  \caption{A section of the autocorrelation function for $^{69}$Ge in iron, using the $^{\mathrm{nat}}$Fe target (run 2). $\chi^2/\nu = 0.70$.  The uncertainties quoted in Table~\ref{tab:omega_results} are taken from maximum and minimum values where $\chi^2 = \nu$.}
  \label{fig:ge69_auto}
\end{figure}

\begin{figure}
  \centering
  \includegraphics[width=0.5\textwidth]{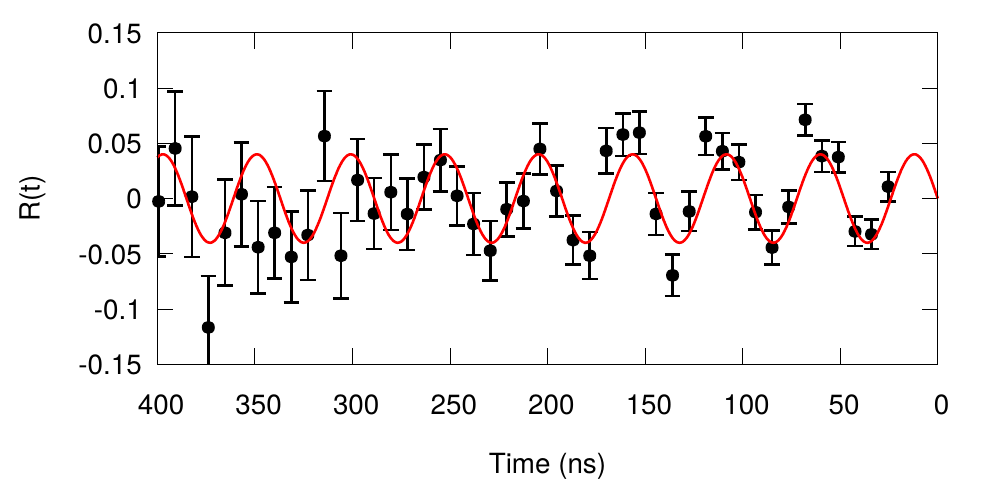}
  \caption{$R(t)$ function for $^{67}$Ge in iron, from the same data set as for $^{69}$Ge in iron in Fig.~\ref{fig:ge69_ratio} and Fig.~\ref{fig:ge69_auto}, using the $^{\mathrm{nat}}$Fe target (run 2). $\chi^2/\nu = 1.18$.}
  \label{fig:ge67_1}
\end{figure}

Two isomers are present in the second data set, studying Ge in iron (run 2) where one iron foil serves as both target and host. The most intense is the $398$-keV, $I^{\pi} = \nicefrac{9}{2}^+$ isomer in $^{69}$Ge with $\tau = 4.05$~$\mu$s, and the other is the corresponding $I^{\pi} = \nicefrac{9}{2}^+$ isomer at $752$ keV in $^{67}$Ge with $\tau = 160$~ns. The 4-$\mu$s isomer allows the behavior of the hyperfine field to be examined over many oscillation periods and is therefore sensitive to any distribution of hyperfine fields that would give rise to beats or attenuation on much longer time scales than the attenuations observed for Cd in gadolinium in Ref.~\cite{Gray2017}.  The $R(t)$ function from $^{69}$Ge is shown in Fig.~\ref{fig:ge69_ratio}. While the long lifetime means that counts are spread in time and thus the precessions are difficult to observe due to statistical fluctuations, forming an autocorrelation function can help establish the frequencies present~\cite{Georgiev2002}.  The autocorrelation function for $^{69}$Ge is shown in Fig.~\ref{fig:ge69_auto}.  For $^{67}$Ge, the $R(t)$ function is shown in Fig.~\ref{fig:ge67_1}.  The Larmor precession frequencies determined from the best fits are given in Table~\ref{tab:omega_results}.

\paragraph*{}
The ratio of the Larmor frequencies of $^{67}$Ge and $^{69}$Ge from Table~\ref{tab:omega_results} is $\omega_L(^{67}\mathrm{Ge})/\omega_L(^{69}\mathrm{Ge}) = 0.869(9)$. Using the same procedure as Ref.~\cite{Lee1991}, this ratio can be used together with $g(^{69}$Ge$) = -0.2224(7)$~\cite{Christiansen1970} to give $g(^{67}$Ge$) = -0.1932(22)$. This new $g$-factor value is in reasonable ($1\sigma$) agreement with the previous measurement of $g(^{67}$Ge$) = -0.1887(26)$~\cite{Lee1991}, however the value of $g(^{67}$Ge$) = -0.210(7)$~\cite{Bertschat1972} is almost $3\sigma$ away.

\paragraph*{}
The $^{69}$Ge measurement allows a precise $\Bhf$ value to be extracted from $\omega_L(^{69}$Ge$) = 74.95(12)$, and $g(^{69}$Ge$) = -0.2224(7)$~\cite{Christiansen1970}. The new measurement at $\approx 6$~K is $\Bhf = +7.036(25)$~T, which compares with $\Bhf = +6.76(5)$~\cite{Lee1991} at $300$~K.%

\begin{figure}
  \centering
  \includegraphics[width=0.5\textwidth]{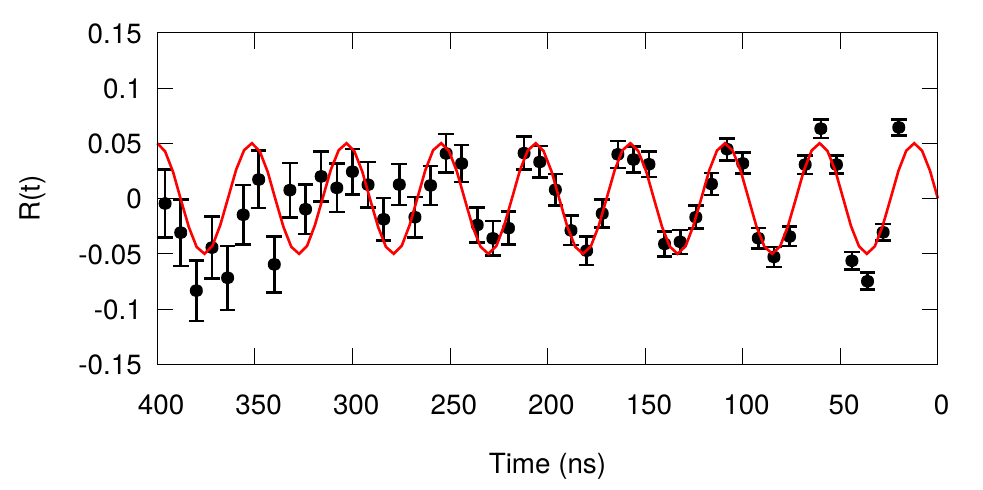}
  \caption{$R(t)$ function for $^{67}$Ge implanted into iron, using the $^{54}$Fe$+^{\mathrm{nat}}$Fe target (run 3). $\chi^2/\nu = 1.20$.}
  \label{fig:gefe}
\end{figure}

\paragraph*{}
In the third run, using the combined $^{54}$Fe and $^{\mathrm{nat}}$Fe target, only $^{67}$Ge was present at a significant level. Similar behaviour to that shown in Fig.~\ref{fig:ge67_1} is observed. The $R(t)$ function and fit are shown in Fig.~\ref{fig:gefe}. Figure~\ref{fig:gefe} may show some weak damping of the $R(t)$ amplitude, however the oscillation is consistent with a constant amplitude from $\approx100$~ns after the prompt ($t=0$). Since the effect is small in magnitude compared to the damping observed for $^{67}$Ge in gadolinium below, it can be considered negligible for the following discussion.
The Larmor precession frequencies obtained from $R(t)$ functions in runs 2 and 3 compare well, as shown in Table~\ref{tab:omega_results}.

\subsection{$^{\mathbf{67}}$Ge in gadolinium}
\begin{figure}
  \centering
  \includegraphics[width=0.5\textwidth]{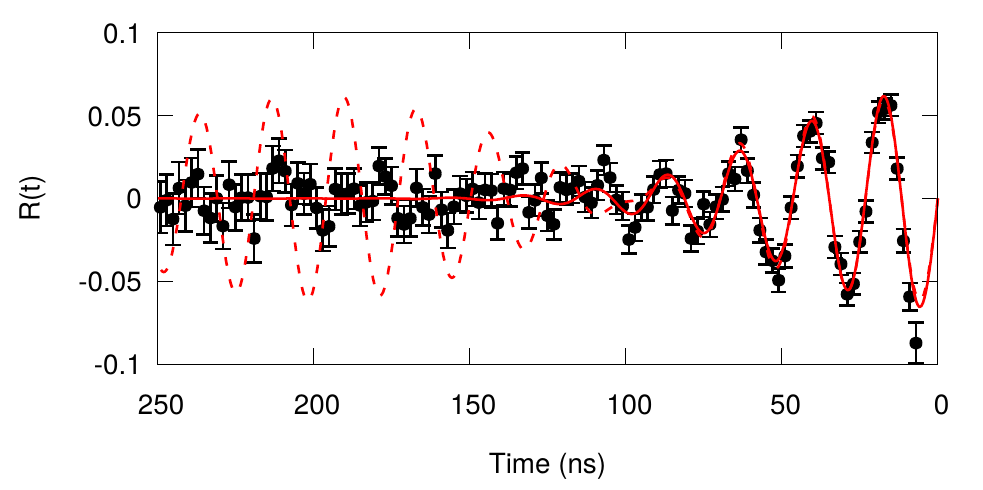}
  \caption{$R(t)$ function for $^{67}$Ge implanted into gadolinium (run 4). Solid fitted line assumes a Gaussian distribution of hyperfine field strengths. This function can be compared to that reported by Raghavan \textit{et al.} (Fig. 12 in Ref.~\cite{Raghavan1985}).  The dotted fitted line includes just two distinct frequencies ($\Bhf = -13.9(2),-15.6(3)$~T), and is fitted to only the first $\approx 100$~ns of data.}
  \label{fig:gegd}
\end{figure}

The $R(t)$ function obtained for the case of $^{67}$Ge in gadolinium (run 4) is shown in Fig.~\ref{fig:gegd}. The attenuation seen here is attributed to a distribution of field strengths as the oscillations in the $R(t)$ function do not come back into phase. If the attenuation was a result of two distinct dominant frequencies ``beating'' in and out of phase, the full amplitude would be restored by $\approx 250$~ns after the prompt. This scenario is shown by the dotted line in Fig.~\ref{fig:gegd}, which is a fit to the first $100$~ns with two distinct frequencies. The beat pattern then implied at $\approx250$~ns is not observed. The solid-line fitted function assumes a Gaussian distribution of field-strengths, as in previous work~\cite{Gray2017,Raghavan1979}.  The free parameters are the mean and FWHM of the field-strength distribution, the initial $R(t)$ amplitude, and a time offset. The fitted parameters are given in Table~\ref{tab:literature}. The $\Bhf$ distribution deduced from Fig.~\ref{fig:gegd} is very similar to that reported in Ref.~\cite{Raghavan1979}, both in mean field-strength and distribution width.

\begin{table}[]
  \centering
  \caption{Comparison of $\omega_L$ values observed in present and previous work for $^{66}$Ga and $^{67,69}$Ge in iron. All values from the present work are determined from fits to the $R(t)$ functions, with the exception of $^{69}$Ge, where the autocorrelation function was used.}
  \label{tab:omega_results}
  \begin{ruledtabular}
  \begin{tabular}{llll}
  Run/    & Impurity  & Temp. & $\omega_L$     \\
  Reference      &           & (K)   & (Mrad s$^{-1}$) \\ \hline
  1              & $^{66}$Ga & 6     & 66.27(33)      \\
  2              & $^{67}$Ge & 6     & 65.1(7)        \\
  3              & $^{67}$Ge & 6     & 65.74(32)      \\
  \cite{Lee1991} & $^{67}$Ge & 300   & 61.1(7)        \\
  2              & $^{69}$Ge & 6     & 74.95(12)     \\
  \cite{Lee1991} & $^{69}$Ge & 300   & 72.0(5)        \\
  \end{tabular}
  \end{ruledtabular}
\end{table}

\begin{table*}
  \centering
  \caption{Static hyperfine fields at dilute Ge and Ga impurities in iron and gadolinium.  FWHM refers to the full width at half maximum of the Gaussian distribution, where fitted.  This is given as both an absolute value, and as a percentage of the mean $\Bhf$ strength.}
  \label{tab:literature}
  \begin{minipage}{\textwidth}
    \renewcommand*\footnoterule{}
    \renewcommand{\thefootnote}{$\alph{footnote}$}
    \begin{tabularx}{\textwidth}{XXXXXXX} \hline \hline
      Impurity   & Host & Temp.       & $\Bhf$        & FWHM     & FWHM    & Ref./Run                            \\
                 &      & (K) & (T)           & (T)      & ($\%$)  &                                     \\\hline
      $^{66}$Ga  & Fe   & 300         & -9.4(5)       &          &         & \cite{Krolas1974}                   \\
      $^{66}$Ga  & Fe   & 4.2         & $\pm 11.0(3)$ &          &         & \cite{Kontani1965}                  \\
      $^{67}$Ge  & Fe   & 300         & +6.7(3)       &          &         & \cite{Raghavan1978}\footnotemark[1] \\
      $^{69}$Ge  & Fe   & 300         & +6.76(5)      &          &         & \cite{Lee1991}                      \\
      $^{67}$Ge  & Gd   & 77          & -14.2(8)      & 2.0(3)   & 14(2)   & \cite{Raghavan1979}\footnotemark[1] \\ \hline
      $^{69}$Ge  & Fe   & 6           & +7.036(25)    &          &         & 2                                   \\
      $^{67}$Ge  & Gd   & 6           & -14.65(17)    & 2.54(10) & 17.3(7) & 4                                   \\ \hline \hline
    \end{tabularx}
    \footnotetext[1]{Adjusted for a subsequent $g$-factor measurement~\cite{Lee1991,Bertschat1972}.}
  \end{minipage}
\end{table*}

\begin{figure}
  \centering
  \includegraphics[width=0.5\textwidth]{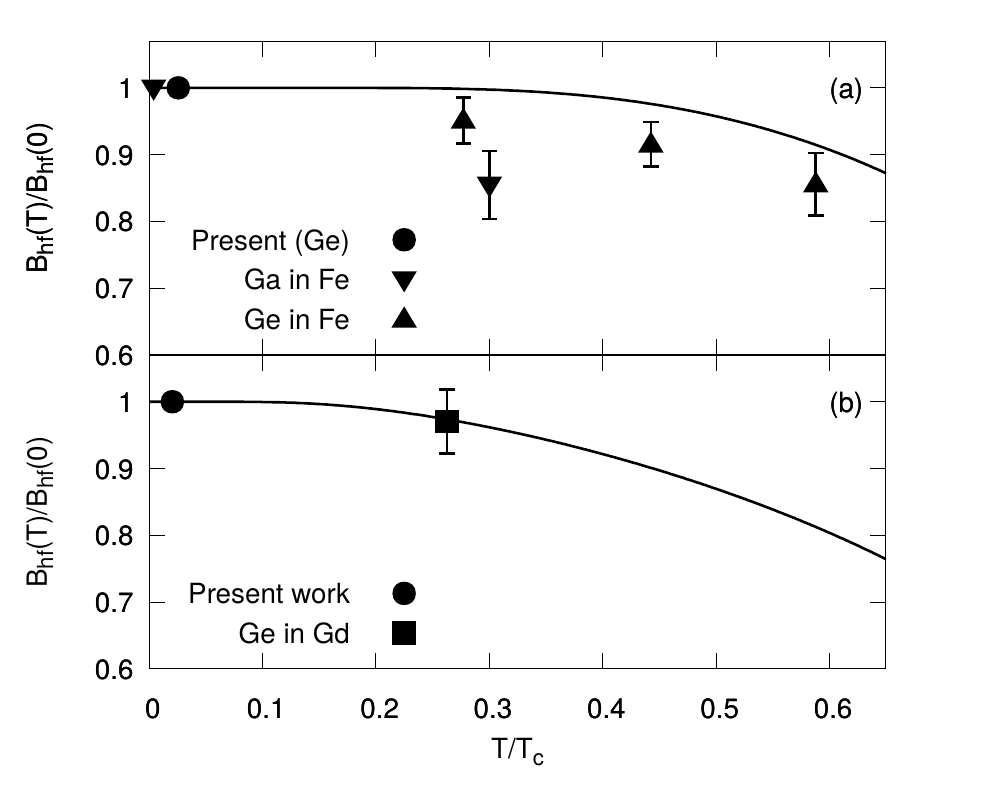}
  \caption{Magnetization of iron (a) and gadolinium (b) hosts as a function of $T/T_c$, where $T_c$ is the Curie temperature. Higher temperature measurements of $\Bhf$ are taken from Refs.~\cite{Kontani1965,Krolas1974,Raghavan1978,Raghavan1979}. Some data points have been displaced along the $x$-axis slightly for display purposes. The solid lines are the Brillouin function, which is an approximation to the magnetization of the host material~\cite{Bozorth1993}.}
  \label{fig:brillouin}
\end{figure}

\section{Discussion}%
\subsection{Comparison with previous measurements}

For both Ge and Ga implanted into iron, an increase in the magnitude of $\Bhf$ is observed at $\approx 6$~K compared to previous measurements at 300~K~\cite{Raghavan1978,Lee1991,Krolas1974}.
Increased $|\Bhf|$ values at $\approx 6$~K are expected since generally $\Bhf$ follows the temperature dependence of the host magnetization, at least approximately (see Fig.~10 in Ref.~\cite{Bernas1973}, or Fig.~5 in Ref.~\cite{Fahlander1979}, for example). Figure~\ref{fig:brillouin} compares the current and previous measurements of the relative field strengths $\Bhf(T)/\Bhf(0)$ to the Brillouin function, which serves as an approximation to the relative magnetization of the host as a function of temperature. The deduced mean $|\Bhf|$ for Ge implanted into gadolinium matches the only previous measurement~\cite{Raghavan1979} very well, as can be seen in Fig.~\ref{fig:brillouin} (b).

Hence the observed $\Bhf$ at 6~K is consistent in all cases with previous work and a temperature dependence that is in accord with expectations. %

\subsection{Amplitude of $\mathbf{R(t)}$ function}%
\label{sec:discussion_amplitude}%
The initial $R(t)$ amplitudes observed in the present work are similar to each other, and to those previously reported for a variety of impurities, with the same or similar mass, in iron, gadolinium, and copper~\cite{Raghavan1985,Lee1991,Mohanta2013}.
\paragraph*{}%
However, the initial $R(t)$ amplitudes are significantly lower than expected for $\sigma/I = 0.35$. In Ref.~\cite{Gray2017}, a reduced amplitude was observed for $^{107}$Cd in gadolinium and it was attributed to a significant fraction of field-free or low-field implantation sites.  If $\sigma/I = 0.35$ is assumed, all cases in the present work give the fraction of field-free implantation sites as $\approx 60\%$, for implantations into both iron and gadolinium.  While this inference might be consistent with previous results for Cd in gadolinium~\cite{Gray2017}, iron has typically been considered a host where essentially all implantations of the $^{67}$Ge and $^{69}$Ge isomers reside on full-field sites~\cite{Raghavan1985,Lee1991}.%
\paragraph*{}%
More specifically, a previous study of the hyperfine fields of Ge$_x$Fe$_{1-x}$ alloys is based on the assumption that the implanted Ge nuclei reside on substitutional (full-field) sites. While this assumption is not specifically evaluated, the results reported support it~\cite{Lee1991} and exclude the possibility of a low-field fraction as high as $60\%$. The implication is then that the reduced amplitude of the $R(t)$ function is a consequence of the nuclear reaction mechanism populating the state.
\paragraph*{}%
Supporting this conclusion, a similar $R(t)$ amplitude to the one observed in the present work was reported in Ref.~\cite{Mohanta2013}.  In that case, $^{66}$Ga was implanted into a non-ferromagnetic Cu foil, using an external field to induce the nuclear precession. In that work, the static hyperfine field is absent and ``field-free'' implantations are irrelevant. The amplitude of $R(t)$ reported in Ref.~\cite{Mohanta2013} corresponds to full alignment from the reaction, and agrees with the amplitudes observed for iron hosts here.  (There is a very weak and slow damping evident in Ref.~\cite{Mohanta2013} that might stem from weak hyperfine interactions or from incomplete background subtraction in the formation of the $R(t)$ function, but it is irrelevant as concerns the initial amplitude of the $R(t)$ function that is of interest here.)
\paragraph*{}%
The $R(t)$ amplitudes measured in the present work indicate that the alignment produced in these isomeric states is much lower than expected from the empirical rule of $\sigma/I \approx 0.35$. If $100\%$ full-field implantation sites are assumed in the present work, alignment parameters of $\sigma/I \approx 0.9$ are necessary to explain the $R(t)$ amplitudes in Figs.~\ref{fig:gafe}, \ref{fig:ge67_1}, and \ref{fig:gefe}.
\paragraph*{}%
This conclusion is not without precedent.  For example Zobel~\textit{et al.}~\cite{Zobel1979} studied high-spin states in $^{69}$Ge, finding generally at high spin that alignments are consistent with $\sigma/I \approx 0.35$; they also observed the 398-keV $9/2^+$ isomer decay to be isotropic.  However it is difficult to disentangle the alignment from the reaction from the subsequent hyperfine interactions in such a measurement. To our knowledge there are few cases where such a separation of reaction mechanism and hyperfine interactions has been reported.
\paragraph*{}%
The general implication of these results is that a measurement of $R(t)$ alone is not a reliable measure of the field-free fraction: either an external-field measurement, or an angular distribution measurement on similar near-by non-isomeric states is needed to verify the nuclear alignment.  The empirical rule $\sigma/I \approx 0.35$ is based on observations of prompt decays.  However, some caution is required if it is assumed equally valid for isomeric states that might sample a different feeding pattern than promptly decaying states.  Note that in Ref.~\cite{Gray2017}, angular distribution measurements on states near and feeding into the isomer in $^{107}$Cd gave an independent measure of the $a_{2}$ value. Unfortunately the level schemes of $^{67,69}$Ge did not allow for a similar approach here.%
\subsection{Comparison of Ge in gadolinium and Cd in gadolinium}
\begin{figure}
  \centering
  \includegraphics[width=0.5\textwidth]{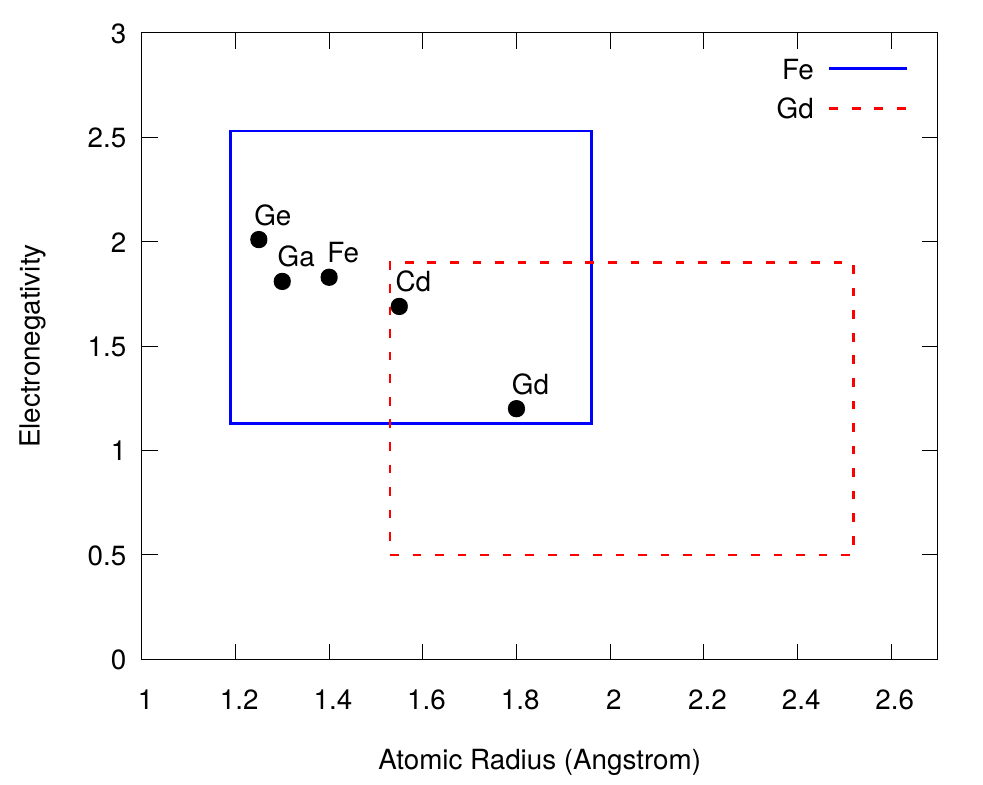}
  \caption{Darken-Gurry plot for relevant alloys. The lines show the empirical solubility limits for Fe (blue continuous line) and Gd (red dotted line) proposed in Ref.~\cite{Sood1978}. Electronegativity data taken from Ref.~\cite{Allred1961}, and atomic radii from Ref.~\cite{Slater1964}.}
  \label{fig:darkengurry}
\end{figure}

The present work indicates that the hyperfine field for Ge in gadolinium behaves more consistently than was observed in the previously studied case of $^{107}$Cd in gadolinium~\cite{Gray2017}.  No variations of average $\Bhf$ strength or $\Bhf$ distribution width were observed over the period of $\approx 30$ hours while the data were being collected.  This is in contrast to the data in Ref.~\cite{Gray2017}, which showed variation of both mean and width of the $\Bhf$ distributions observed on a timescale of $\approx 10$ hours. One important difference is that the beam current was much higher during the Cd measurements (to mimic earlier the IPAD measurement~\cite{Regan1995}).%
\paragraph*{}%
In general, the simplicity or complexity of the $\Bhf$ distribution after implantation is related to how well the system alloys.  Thus, we consider the empirical rules that guide whether two metals will alloy or not: (i) the matching of the atomic radius, and (ii), the matching of the electronegativity~\cite{Sood1978}.%
\paragraph*{}%
Fig.~\ref{fig:darkengurry} shows the Darken-Gurry plot~\cite{Sood1978} for the impurities and hosts considered.  Note that Ge and Ga are well matched with iron for both atomic radius, and electronegativity.  However, Ge, Ga, and Cd are all quite mismatched with the electronegativity of gadolinium. Hence, the distribution of $|\Bhf|$ fields observed in the gadolinium host can be interpreted to be a result of the system not alloying easily. However, it seems that the difference in the behaviours of Ge and Cd implanted into gadolinium hosts is not a result of the electronegativity mismatch, since the mismatch is similar for both Ge and Cd. Additionally, the mismatch in atomic radius is greater for Ge than for Cd.%
\paragraph*{}%
These observations support our suggestion that the time-dependent variation in the fields encountered for the case of Cd in gadolinium may have been caused by accumulating radiation damage, exacerbated by the high beam intensity used. In the case of Ref.~\cite{Gray2017}, a TDPAD measurement could have been achieved with a lower beam dose.%
\subsection{New $\mathbf{g}$-factor measurements}
 The new $g$-factor measurements for $^{67}$Ge and $^{66}$Ga using iron as a ferromagnetic host are in agreement with previous work~\cite{Lee1991,Filevich1978}.%
\paragraph*{}%
If the previous $\Bhf = -14.2(8)$~T~\cite{Raghavan1985,Raghavan1979} for Ge in gadolinium is adjusted for a $6$~K temperature using the Brillouin function, a field strength of $\Bhf = -14.6(9)$~T is obtained.  This field can be used with the $R(t)$ function from run 4 shown in Fig.~\ref{fig:gegd} to obtain $g(\nicefrac{9}{2}^+, ^{67}\mathrm{Ge}) = -0.188(11)$, in excellent agreement with the independently determined value of $g = -0.1932(22)$ (run 2), as well as the previous literature value of $g = -0.1887(27)$~\cite{Lee1991}.  This consistency and the fact that the uncertainty stems primarily from the uncertainty in the literature value of the hyperfine field strength demonstrates that gadolinium hosts can be effective for TDPAD $g$-factor measurements.%
\paragraph*{}%
In the $g$-factor measurement of Filevich~\textit{et al.}~\cite{Filevich1978}, the spin of the $7^-$ isomeric state in $^{66}$Ga was incorrectly assigned as $I^\pi = 6^-$. The proposed configuration of the isomer is $[\pi f_{5/2} \otimes \nu g_{9/2}]_{I^-}$. The present work on $^{67}$Ge suggests an empirical value of $g = -0.19$ for the $g_{\nicefrac{9}{2}}$ neutron, while the ground state of $^{69}$As suggests an empirical value for the $f_{\nicefrac{5}{2}}$ proton of $g = +0.649$~\cite{Golovko2005}. Adopting these empirical $g$ factors and coupling the spins to $7^-$ gives $g = +0.11$, in much better agreement with the experimental $g = +0.126(4)$ than coupling to $6^-$, which gives $g = +0.07$.
\section{Conclusion}
Gadolinium and iron have been examined as ferromagnetic hosts for TDPAD measurements using isomers in $^{66}$Ga, $^{67}$Ge, and $^{69}$Ge as probes. New values for $g$ factors of excited states in $^{67}$Ge and $^{66}$Ga have been determined, which are in agreement with previous work.  Additionally, the static hyperfine field strength, $\Bhf$, has been measured at $\approx 6$~K for Ge implanted into iron and gadolinium hosts.  These results are in accord with expectations of the temperature-dependent nature of $\Bhf$. Our $g$-factor results support the configuration assignment of $[\pi f_{5/2} \otimes \nu g_{9/2}]_{7^-}$ to the isomeric state in $^{66}$Ga.%
\paragraph*{}%
The initial amplitude of the $R(t)$ function depends on both the spin-alignment of the isomeric state and the fraction of implanted nuclei on field-free sites. Comparisons of present and previous results indicate that care must be taken when inferring the fraction of field-free implantations from the $R(t)$ amplitude; an independent measurement of the alignment or of the $R(t)$ amplitude applicable to implantation onto 100\% full-field sites is needed.
\paragraph*{}%
The results for Ge after implantation into gadolinium support the conclusion that the time-dependent variations in $\Bhf$ reported in Ref.~\cite{Gray2017} were largely a result of the high beam intensities used. We have shown that a stable distribution of $\Bhf$ can be obtained independently in different laboratories, indicating that gadolinium can serve as a suitable host for precision $g$-factor measurements.
\begin{acknowledgements}
  The authors are grateful to the academic and technical staff of the Department of Nuclear Physics and the Heavy Ion Accelerator Facility (Australian National University). Thanks to A. Akber, T. Palazzo, and T. Tornyi for their help with data collection. T.J.G, B.J.C, J.T.H.D, M.S.M.G, and B.P.M acknowledge the support of the Australian Government Research Training Program.  This research was supported in part by Australian Research Council grant number DP170701673.  Support for the ANU Heavy Ion Accelerator Facility operations through the Australian National Collaborative Research Infrastructure Strategy (NCRIS) program is acknowledged.
\end{acknowledgements}

\bibliography{GeGa_in_fegd}

%apsrev4-2.bst 2019-01-14 (MD) hand-edited version of apsrev4-1.bst
%Control: key (0)
%Control: author (8) initials jnrlst
%Control: editor formatted (1) identically to author
%Control: production of article title (0) allowed
%Control: page (0) single
%Control: year (1) truncated
%Control: production of eprint (0) enabled
\begin{thebibliography}{43}%
\makeatletter
\providecommand \@ifxundefined [1]{%
 \@ifx{#1\undefined}
}%
\providecommand \@ifnum [1]{%
 \ifnum #1\expandafter \@firstoftwo
 \else \expandafter \@secondoftwo
 \fi
}%
\providecommand \@ifx [1]{%
 \ifx #1\expandafter \@firstoftwo
 \else \expandafter \@secondoftwo
 \fi
}%
\providecommand \natexlab [1]{#1}%
\providecommand \enquote  [1]{``#1''}%
\providecommand \bibnamefont  [1]{#1}%
\providecommand \bibfnamefont [1]{#1}%
\providecommand \citenamefont [1]{#1}%
\providecommand \href@noop [0]{\@secondoftwo}%
\providecommand \href [0]{\begingroup \@sanitize@url \@href}%
\providecommand \@href[1]{\@@startlink{#1}\@@href}%
\providecommand \@@href[1]{\endgroup#1\@@endlink}%
\providecommand \@sanitize@url [0]{\catcode `\\12\catcode `\$12\catcode
  `\&12\catcode `\#12\catcode `\^12\catcode `\_12\catcode `\%12\relax}%
\providecommand \@@startlink[1]{}%
\providecommand \@@endlink[0]{}%
\providecommand \url  [0]{\begingroup\@sanitize@url \@url }%
\providecommand \@url [1]{\endgroup\@href {#1}{\urlprefix }}%
\providecommand \urlprefix  [0]{URL }%
\providecommand \Eprint [0]{\href }%
\providecommand \doibase [0]{https://doi.org/}%
\providecommand \selectlanguage [0]{\@gobble}%
\providecommand \bibinfo  [0]{\@secondoftwo}%
\providecommand \bibfield  [0]{\@secondoftwo}%
\providecommand \translation [1]{[#1]}%
\providecommand \BibitemOpen [0]{}%
\providecommand \bibitemStop [0]{}%
\providecommand \bibitemNoStop [0]{.\EOS\space}%
\providecommand \EOS [0]{\spacefactor3000\relax}%
\providecommand \BibitemShut  [1]{\csname bibitem#1\endcsname}%
\let\auto@bib@innerbib\@empty
%</preamble>
\bibitem [{\citenamefont {Brown}\ \emph {et~al.}(2005)\citenamefont {Brown},
  \citenamefont {Stone}, \citenamefont {Stone}, \citenamefont {Towner},\ and\
  \citenamefont {Hjorth-Jensen}}]{Brown2004}%
  \BibitemOpen
  \bibfield  {author} {\bibinfo {author} {\bibfnamefont {B.~A.}\ \bibnamefont
  {Brown}}, \bibinfo {author} {\bibfnamefont {N.~J.}\ \bibnamefont {Stone}},
  \bibinfo {author} {\bibfnamefont {J.~R.}\ \bibnamefont {Stone}}, \bibinfo
  {author} {\bibfnamefont {I.~S.}\ \bibnamefont {Towner}},\ and\ \bibinfo
  {author} {\bibfnamefont {M.}~\bibnamefont {Hjorth-Jensen}},\ }\bibfield
  {title} {\bibinfo {title} {Magnetic moments of the ${2}_{1}^{+}$ states
  around $^{132}\mathrm{Sn}$},\ }\href@noop {} {\bibfield  {journal} {\bibinfo
  {journal} {Phys. Rev. C}\ }\textbf {\bibinfo {volume} {71}},\ \bibinfo
  {pages} {044317} (\bibinfo {year} {2005})}\BibitemShut {NoStop}%
\bibitem [{\citenamefont {Mertzimekis}\ \emph {et~al.}(2003)\citenamefont
  {Mertzimekis}, \citenamefont {Stuchbery}, \citenamefont {Benczer-Koller},\
  and\ \citenamefont {Taylor}}]{Mertzimekis2003}%
  \BibitemOpen
  \bibfield  {author} {\bibinfo {author} {\bibfnamefont {T.~J.}\ \bibnamefont
  {Mertzimekis}}, \bibinfo {author} {\bibfnamefont {A.~E.}\ \bibnamefont
  {Stuchbery}}, \bibinfo {author} {\bibfnamefont {N.}~\bibnamefont
  {Benczer-Koller}},\ and\ \bibinfo {author} {\bibfnamefont {M.~J.}\
  \bibnamefont {Taylor}},\ }\bibfield  {title} {\bibinfo {title} {Systematics
  of first ${2}^{+}$ state $g$ factors around mass 80},\ }\href@noop {}
  {\bibfield  {journal} {\bibinfo  {journal} {Phys. Rev. C}\ }\textbf {\bibinfo
  {volume} {68}},\ \bibinfo {pages} {054304} (\bibinfo {year}
  {2003})}\BibitemShut {NoStop}%
\bibitem [{\citenamefont {Kusakari}\ \emph {et~al.}(1984)\citenamefont
  {Kusakari}, \citenamefont {Sugawara}, \citenamefont {Fujioka}, \citenamefont
  {Kawamura}, \citenamefont {Hayashibe}, \citenamefont {Iura}, \citenamefont
  {Sakai},\ and\ \citenamefont {Ishimatsu}}]{Kusakari1984}%
  \BibitemOpen
  \bibfield  {author} {\bibinfo {author} {\bibfnamefont {H.}~\bibnamefont
  {Kusakari}}, \bibinfo {author} {\bibfnamefont {M.}~\bibnamefont {Sugawara}},
  \bibinfo {author} {\bibfnamefont {M.}~\bibnamefont {Fujioka}}, \bibinfo
  {author} {\bibfnamefont {N.}~\bibnamefont {Kawamura}}, \bibinfo {author}
  {\bibfnamefont {S.}~\bibnamefont {Hayashibe}}, \bibinfo {author}
  {\bibfnamefont {K.}~\bibnamefont {Iura}}, \bibinfo {author} {\bibfnamefont
  {F.}~\bibnamefont {Sakai}},\ and\ \bibinfo {author} {\bibfnamefont
  {T.}~\bibnamefont {Ishimatsu}},\ }\bibfield  {title} {\bibinfo {title}
  {Nuclear $g$ factor of the 2972 kev isomeric state in $^{130}\mathrm{Xe}$},\
  }\href@noop {} {\bibfield  {journal} {\bibinfo  {journal} {Phys. Rev. C}\
  }\textbf {\bibinfo {volume} {30}},\ \bibinfo {pages} {820} (\bibinfo {year}
  {1984})}\BibitemShut {NoStop}%
\bibitem [{\citenamefont {Ingwersen}\ \emph {et~al.}(1975)\citenamefont
  {Ingwersen}, \citenamefont {Klinger}, \citenamefont {Schatz},\ and\
  \citenamefont {Witthuhn}}]{Ingwersen1975}%
  \BibitemOpen
  \bibfield  {author} {\bibinfo {author} {\bibfnamefont {H.}~\bibnamefont
  {Ingwersen}}, \bibinfo {author} {\bibfnamefont {W.}~\bibnamefont {Klinger}},
  \bibinfo {author} {\bibfnamefont {G.}~\bibnamefont {Schatz}},\ and\ \bibinfo
  {author} {\bibfnamefont {W.}~\bibnamefont {Witthuhn}},\ }\bibfield  {title}
  {\bibinfo {title} {Nuclear $g$ factors of the
  ${\frac{21}{2}}^{\ensuremath{-}}$ and the ${\frac{29}{2}}^{+}$ states in
  $^{211}\mathrm{At}$},\ }\href@noop {} {\bibfield  {journal} {\bibinfo
  {journal} {Phys. Rev. C}\ }\textbf {\bibinfo {volume} {11}},\ \bibinfo
  {pages} {243} (\bibinfo {year} {1975})}\BibitemShut {NoStop}%
\bibitem [{\citenamefont {M\"ucher}\ \emph {et~al.}(2009)\citenamefont
  {M\"ucher}, \citenamefont {G\"urdal}, \citenamefont {Speidel}, \citenamefont
  {Kumbartzki}, \citenamefont {Benczer-Koller}, \citenamefont {Robinson},
  \citenamefont {Sharon}, \citenamefont {Zamick}, \citenamefont {Lisetskiy},
  \citenamefont {Casperson}, \citenamefont {Heinz}, \citenamefont {Krieger},
  \citenamefont {Leske}, \citenamefont {Maier-Komor}, \citenamefont {Werner},
  \citenamefont {Williams},\ and\ \citenamefont {Winkler}}]{Mucher2009}%
  \BibitemOpen
  \bibfield  {author} {\bibinfo {author} {\bibfnamefont {D.}~\bibnamefont
  {M\"ucher}}, \bibinfo {author} {\bibfnamefont {G.}~\bibnamefont {G\"urdal}},
  \bibinfo {author} {\bibfnamefont {K.-H.}\ \bibnamefont {Speidel}}, \bibinfo
  {author} {\bibfnamefont {G.~J.}\ \bibnamefont {Kumbartzki}}, \bibinfo
  {author} {\bibfnamefont {N.}~\bibnamefont {Benczer-Koller}}, \bibinfo
  {author} {\bibfnamefont {S.~J.~Q.}\ \bibnamefont {Robinson}}, \bibinfo
  {author} {\bibfnamefont {Y.~Y.}\ \bibnamefont {Sharon}}, \bibinfo {author}
  {\bibfnamefont {L.}~\bibnamefont {Zamick}}, \bibinfo {author} {\bibfnamefont
  {A.~F.}\ \bibnamefont {Lisetskiy}}, \bibinfo {author} {\bibfnamefont {R.~J.}\
  \bibnamefont {Casperson}}, \bibinfo {author} {\bibfnamefont {A.}~\bibnamefont
  {Heinz}}, \bibinfo {author} {\bibfnamefont {B.}~\bibnamefont {Krieger}},
  \bibinfo {author} {\bibfnamefont {J.}~\bibnamefont {Leske}}, \bibinfo
  {author} {\bibfnamefont {P.}~\bibnamefont {Maier-Komor}}, \bibinfo {author}
  {\bibfnamefont {V.}~\bibnamefont {Werner}}, \bibinfo {author} {\bibfnamefont
  {E.}~\bibnamefont {Williams}},\ and\ \bibinfo {author} {\bibfnamefont
  {R.}~\bibnamefont {Winkler}},\ }\bibfield  {title} {\bibinfo {title} {Nuclear
  structure studies of $^{70}\mathrm{Zn}$ from $g$-factor and lifetime
  measurements},\ }\href@noop {} {\bibfield  {journal} {\bibinfo  {journal}
  {Phys. Rev. C}\ }\textbf {\bibinfo {volume} {79}},\ \bibinfo {pages} {054310}
  (\bibinfo {year} {2009})}\BibitemShut {NoStop}%
\bibitem [{\citenamefont {Morinaga}\ and\ \citenamefont
  {Yamazaki}(1976)}]{Morinaga1976}%
  \BibitemOpen
  \bibfield  {author} {\bibinfo {author} {\bibfnamefont {H.}~\bibnamefont
  {Morinaga}}\ and\ \bibinfo {author} {\bibfnamefont {T.}~\bibnamefont
  {Yamazaki}},\ }\bibinfo {title} {In-beam gamma-ray spectroscopy}\ (\bibinfo
  {publisher} {North-Holland Publishing Company},\ \bibinfo {year} {1976})\
  Chap.~\bibinfo {chapter} {9}, pp.\ \bibinfo {pages} {424 -- 479}\BibitemShut
  {NoStop}%
\bibitem [{\citenamefont {Raghavan}\ and\ \citenamefont
  {Raghavan}(1985)}]{Raghavan1985}%
  \BibitemOpen
  \bibfield  {author} {\bibinfo {author} {\bibfnamefont {P.}~\bibnamefont
  {Raghavan}}\ and\ \bibinfo {author} {\bibfnamefont {R.~S.}\ \bibnamefont
  {Raghavan}},\ }\bibfield  {title} {\bibinfo {title} {Hyperfine interaction
  studies with pulsed heavy-ion beams},\ }\href@noop {} {\bibfield  {journal}
  {\bibinfo  {journal} {Hyperfine Interactions}\ }\textbf {\bibinfo {volume}
  {26}},\ \bibinfo {pages} {855} (\bibinfo {year} {1985})}\BibitemShut
  {NoStop}%
\bibitem [{\citenamefont {Stuchbery}\ \emph {et~al.}(2020)\citenamefont
  {Stuchbery}, \citenamefont {Harding}, \citenamefont {Weisser},\ and\
  \citenamefont {Lobanov}}]{Stuchbery2019}%
  \BibitemOpen
  \bibfield  {author} {\bibinfo {author} {\bibfnamefont {A.~E.}\ \bibnamefont
  {Stuchbery}}, \bibinfo {author} {\bibfnamefont {A.~B.}\ \bibnamefont
  {Harding}}, \bibinfo {author} {\bibfnamefont {D.~C.}\ \bibnamefont
  {Weisser}},\ and\ \bibinfo {author} {\bibfnamefont {N.~R.}\ \bibnamefont
  {Lobanov}},\ }\bibfield  {title} {\bibinfo {title} {Apparatus for in-beam
  hyperfine interactions and g-factor measurements: Design and operation},\
  }\href@noop {} {\bibfield  {journal} {\bibinfo  {journal} {Nuclear
  Instruments and Methods in Physics Research Section A: Accelerators,
  Spectrometers, Detectors and Associated Equipment}\ }\textbf {\bibinfo
  {volume} {951}},\ \bibinfo {pages} {162985} (\bibinfo {year}
  {2020})}\BibitemShut {NoStop}%
\bibitem [{\citenamefont {Fahlander}\ \emph
  {et~al.}(1979{\natexlab{a}})\citenamefont {Fahlander}, \citenamefont
  {Johansson},\ and\ \citenamefont {Possnert}}]{Fahlander1979_2}%
  \BibitemOpen
  \bibfield  {author} {\bibinfo {author} {\bibfnamefont {C.}~\bibnamefont
  {Fahlander}}, \bibinfo {author} {\bibfnamefont {K.}~\bibnamefont
  {Johansson}},\ and\ \bibinfo {author} {\bibfnamefont {G.}~\bibnamefont
  {Possnert}},\ }\bibfield  {title} {\bibinfo {title} {The magnetic moment of
  the excited 5/2- state in the 57fe},\ }\href@noop {} {\bibfield  {journal}
  {\bibinfo  {journal} {Physica Scripta}\ }\textbf {\bibinfo {volume} {20}},\
  \bibinfo {pages} {163} (\bibinfo {year} {1979}{\natexlab{a}})}\BibitemShut
  {NoStop}%
\bibitem [{\citenamefont {Rafailovich}\ \emph {et~al.}(1983)\citenamefont
  {Rafailovich}, \citenamefont {Dafni}, \citenamefont {Brennan},\ and\
  \citenamefont {Sprouse}}]{Rafailovich1983}%
  \BibitemOpen
  \bibfield  {author} {\bibinfo {author} {\bibfnamefont {M.~H.}\ \bibnamefont
  {Rafailovich}}, \bibinfo {author} {\bibfnamefont {E.}~\bibnamefont {Dafni}},
  \bibinfo {author} {\bibfnamefont {J.~M.}\ \bibnamefont {Brennan}},\ and\
  \bibinfo {author} {\bibfnamefont {G.~D.}\ \bibnamefont {Sprouse}},\
  }\bibfield  {title} {\bibinfo {title} {Magnetic moment of the
  $^{54}\mathrm{Fe}$(${10}^{+}$) level},\ }\href@noop {} {\bibfield  {journal}
  {\bibinfo  {journal} {Phys. Rev. C}\ }\textbf {\bibinfo {volume} {27}},\
  \bibinfo {pages} {602} (\bibinfo {year} {1983})}\BibitemShut {NoStop}%
\bibitem [{\citenamefont {Stuchbery}\ \emph {et~al.}(2018)\citenamefont
  {Stuchbery}, \citenamefont {McCormick}, \citenamefont {Gray},\ and\
  \citenamefont {Coombes}}]{Stuchbery2018}%
  \BibitemOpen
  \bibfield  {author} {\bibinfo {author} {\bibfnamefont {A.~E.}\ \bibnamefont
  {Stuchbery}}, \bibinfo {author} {\bibfnamefont {B.~P.}\ \bibnamefont
  {McCormick}}, \bibinfo {author} {\bibfnamefont {T.~J.}\ \bibnamefont
  {Gray}},\ and\ \bibinfo {author} {\bibfnamefont {B.~J.}\ \bibnamefont
  {Coombes}},\ }\bibfield  {title} {\bibinfo {title} {Pushing the limits of
  excited-state g-factor measurements},\ }\href@noop {} {\bibfield  {journal}
  {\bibinfo  {journal} {EPJ Web Conf.}\ }\textbf {\bibinfo {volume} {178}},\
  \bibinfo {pages} {02005} (\bibinfo {year} {2018})}\BibitemShut {NoStop}%
\bibitem [{\citenamefont {Hensler}\ \emph {et~al.}(1971)\citenamefont
  {Hensler}, \citenamefont {Tape}, \citenamefont {Benczer-Koller},\ and\
  \citenamefont {MacDonald}}]{Hensler1971}%
  \BibitemOpen
  \bibfield  {author} {\bibinfo {author} {\bibfnamefont {R.}~\bibnamefont
  {Hensler}}, \bibinfo {author} {\bibfnamefont {J.~W.}\ \bibnamefont {Tape}},
  \bibinfo {author} {\bibfnamefont {N.}~\bibnamefont {Benczer-Koller}},\ and\
  \bibinfo {author} {\bibfnamefont {J.~R.}\ \bibnamefont {MacDonald}},\
  }\bibfield  {title} {\bibinfo {title} {Mean lifetime and magnetic moment of
  the 2.95-mev (${6}^{+}$) state of $^{54}\mathrm{Fe}$},\ }\href@noop {}
  {\bibfield  {journal} {\bibinfo  {journal} {Phys. Rev. Lett.}\ }\textbf
  {\bibinfo {volume} {27}},\ \bibinfo {pages} {1587} (\bibinfo {year}
  {1971})}\BibitemShut {NoStop}%
\bibitem [{\citenamefont {Mohn}(2000)}]{Mohn2000}%
  \BibitemOpen
  \bibfield  {author} {\bibinfo {author} {\bibfnamefont {P.}~\bibnamefont
  {Mohn}},\ }\bibfield  {title} {\bibinfo {title} {Theoretical aspects of
  hyperfine interactions},\ }\href@noop {} {\bibfield  {journal} {\bibinfo
  {journal} {Hyperfine Interactions}\ }\textbf {\bibinfo {volume} {128}},\
  \bibinfo {pages} {67} (\bibinfo {year} {2000})}\BibitemShut {NoStop}%
\bibitem [{\citenamefont {Gray}\ \emph {et~al.}(2017)\citenamefont {Gray},
  \citenamefont {Stuchbery}, \citenamefont {Reed}, \citenamefont {Akber},
  \citenamefont {Coombes}, \citenamefont {Dowie}, \citenamefont {Eriksen},
  \citenamefont {Gerathy}, \citenamefont {Kib\'edi}, \citenamefont {Lane},
  \citenamefont {Mitchell}, \citenamefont {Palazzo},\ and\ \citenamefont
  {Tornyi}}]{Gray2017}%
  \BibitemOpen
  \bibfield  {author} {\bibinfo {author} {\bibfnamefont {T.~J.}\ \bibnamefont
  {Gray}}, \bibinfo {author} {\bibfnamefont {A.~E.}\ \bibnamefont {Stuchbery}},
  \bibinfo {author} {\bibfnamefont {M.~W.}\ \bibnamefont {Reed}}, \bibinfo
  {author} {\bibfnamefont {A.}~\bibnamefont {Akber}}, \bibinfo {author}
  {\bibfnamefont {B.~J.}\ \bibnamefont {Coombes}}, \bibinfo {author}
  {\bibfnamefont {J.~T.~H.}\ \bibnamefont {Dowie}}, \bibinfo {author}
  {\bibfnamefont {T.~K.}\ \bibnamefont {Eriksen}}, \bibinfo {author}
  {\bibfnamefont {M.~S.~M.}\ \bibnamefont {Gerathy}}, \bibinfo {author}
  {\bibfnamefont {T.}~\bibnamefont {Kib\'edi}}, \bibinfo {author}
  {\bibfnamefont {G.~J.}\ \bibnamefont {Lane}}, \bibinfo {author}
  {\bibfnamefont {A.~J.}\ \bibnamefont {Mitchell}}, \bibinfo {author}
  {\bibfnamefont {T.}~\bibnamefont {Palazzo}},\ and\ \bibinfo {author}
  {\bibfnamefont {T.}~\bibnamefont {Tornyi}},\ }\bibfield  {title} {\bibinfo
  {title} {Perturbed angular distributions with ${\mathrm{labr}}_{3}$
  detectors: The $g$ factor of the first ${10}^{+}$ state in
  $^{110}\mathrm{Cd}$ reexamined},\ }\href@noop {} {\bibfield  {journal}
  {\bibinfo  {journal} {Phys. Rev. C}\ }\textbf {\bibinfo {volume} {96}},\
  \bibinfo {pages} {054332} (\bibinfo {year} {2017})}\BibitemShut {NoStop}%
\bibitem [{\citenamefont {Regan}\ \emph {et~al.}(1995)\citenamefont {Regan},
  \citenamefont {Stuchbery},\ and\ \citenamefont {Anderssen}}]{Regan1995}%
  \BibitemOpen
  \bibfield  {author} {\bibinfo {author} {\bibfnamefont {P.~H.}\ \bibnamefont
  {Regan}}, \bibinfo {author} {\bibfnamefont {A.~E.}\ \bibnamefont
  {Stuchbery}},\ and\ \bibinfo {author} {\bibfnamefont {S.~S.}\ \bibnamefont
  {Anderssen}},\ }\bibfield  {title} {\bibinfo {title} {Measurement of the
  g-factor of the yrast 10+ state in 110cd},\ }\href@noop {} {\bibfield
  {journal} {\bibinfo  {journal} {Nuclear Physics A}\ }\textbf {\bibinfo
  {volume} {591}},\ \bibinfo {pages} {533 } (\bibinfo {year}
  {1995})}\BibitemShut {NoStop}%
\bibitem [{ENS()}]{ENSDF}%
  \BibitemOpen
  \href@noop {} {\bibinfo {title} {{Evaluated Nuclear Structure Data File
  (ENSDF)}}},\ \bibinfo {howpublished}
  {\url{http://www.nndc.bnl.gov/ensdf/}}\BibitemShut {NoStop}%
\bibitem [{\citenamefont {Forker}\ and\ \citenamefont
  {Hammesfahr}(1973)}]{Forker1973}%
  \BibitemOpen
  \bibfield  {author} {\bibinfo {author} {\bibfnamefont {M.}~\bibnamefont
  {Forker}}\ and\ \bibinfo {author} {\bibfnamefont {A.}~\bibnamefont
  {Hammesfahr}},\ }\bibfield  {title} {\bibinfo {title} {{The magnetic
  hyperfine field at Cd nuclei in the rare earth ferromagnets Gd, Tb, Dy, Ho,
  Er and Tm}},\ }\href@noop {} {\bibfield  {journal} {\bibinfo  {journal}
  {Zeitschrift f{{\"{u}}}r Physik}\ }\textbf {\bibinfo {volume} {263}},\
  \bibinfo {pages} {33} (\bibinfo {year} {1973})}\BibitemShut {NoStop}%
\bibitem [{\citenamefont {Lee}\ \emph {et~al.}(1991)\citenamefont {Lee},
  \citenamefont {Raghavan},\ and\ \citenamefont {Raghavan}}]{Lee1991}%
  \BibitemOpen
  \bibfield  {author} {\bibinfo {author} {\bibfnamefont {C.~S.}\ \bibnamefont
  {Lee}}, \bibinfo {author} {\bibfnamefont {P.}~\bibnamefont {Raghavan}},\ and\
  \bibinfo {author} {\bibfnamefont {R.~S.}\ \bibnamefont {Raghavan}},\
  }\bibfield  {title} {\bibinfo {title} {In-beam study of spin density
  oscillations in ferromagnetic fe-based alloys using 67,69ge isomers},\
  }\href@noop {} {\bibfield  {journal} {\bibinfo  {journal} {Nuclear
  Instruments and Methods in Physics Research Section B: Beam Interactions with
  Materials and Atoms}\ }\textbf {\bibinfo {volume} {56-57}},\ \bibinfo {pages}
  {851 } (\bibinfo {year} {1991})}\BibitemShut {NoStop}%
\bibitem [{\citenamefont {Raghavan}\ \emph {et~al.}(1978)\citenamefont
  {Raghavan}, \citenamefont {Senba},\ and\ \citenamefont
  {Raghavan}}]{Raghavan1978}%
  \BibitemOpen
  \bibfield  {author} {\bibinfo {author} {\bibfnamefont {P.}~\bibnamefont
  {Raghavan}}, \bibinfo {author} {\bibfnamefont {M.}~\bibnamefont {Senba}},\
  and\ \bibinfo {author} {\bibfnamefont {R.~S.}\ \bibnamefont {Raghavan}},\
  }\bibfield  {title} {\bibinfo {title} {Hyperfine magnetic fields at67ge in
  fe, co and ni},\ }\href@noop {} {\bibfield  {journal} {\bibinfo  {journal}
  {Hyperfine Interactions}\ }\textbf {\bibinfo {volume} {4}},\ \bibinfo {pages}
  {330} (\bibinfo {year} {1978})}\BibitemShut {NoStop}%
\bibitem [{\citenamefont {Raghavan}\ \emph {et~al.}(1979)\citenamefont
  {Raghavan}, \citenamefont {Senba},\ and\ \citenamefont
  {Raghavan}}]{Raghavan1979}%
  \BibitemOpen
  \bibfield  {author} {\bibinfo {author} {\bibfnamefont {P.}~\bibnamefont
  {Raghavan}}, \bibinfo {author} {\bibfnamefont {M.}~\bibnamefont {Senba}},\
  and\ \bibinfo {author} {\bibfnamefont {R.}~\bibnamefont {Raghavan}},\
  }\bibfield  {title} {\bibinfo {title} {Magnetic hyperfine fields at ga, ge,
  and as impurities in ferromagnetic gadolinium},\ }\href@noop {} {\bibfield
  {journal} {\bibinfo  {journal} {Bulletin of the American Physical Society}\
  }\textbf {\bibinfo {volume} {24}},\ \bibinfo {pages} {643} (\bibinfo {year}
  {1979})}\BibitemShut {NoStop}%
\bibitem [{\citenamefont {Filevich}\ \emph {et~al.}(1978)\citenamefont
  {Filevich}, \citenamefont {Ceballos}, \citenamefont {Mariscotti},
  \citenamefont {Thieberger},\ and\ \citenamefont {Mateosian}}]{Filevich1978}%
  \BibitemOpen
  \bibfield  {author} {\bibinfo {author} {\bibfnamefont {A.}~\bibnamefont
  {Filevich}}, \bibinfo {author} {\bibfnamefont {A.}~\bibnamefont {Ceballos}},
  \bibinfo {author} {\bibfnamefont {M.~A.~J.}\ \bibnamefont {Mariscotti}},
  \bibinfo {author} {\bibfnamefont {P.}~\bibnamefont {Thieberger}},\ and\
  \bibinfo {author} {\bibfnamefont {E.~D.}\ \bibnamefont {Mateosian}},\
  }\bibfield  {title} {\bibinfo {title} {Lifetimes and g-factors of the 6$^-$
  and 7$^-$ isomers in $^{66}$ga and $^{68}$ga},\ }\href@noop {} {\bibfield
  {journal} {\bibinfo  {journal} {Nuclear Physics A}\ }\textbf {\bibinfo
  {volume} {295}},\ \bibinfo {pages} {513 } (\bibinfo {year}
  {1978})}\BibitemShut {NoStop}%
\bibitem [{\citenamefont {Bozorth}(1993)}]{Bozorth1993}%
  \BibitemOpen
  \bibfield  {author} {\bibinfo {author} {\bibfnamefont {R.~M.}\ \bibnamefont
  {Bozorth}},\ }\bibinfo {title} {{Ferromagnetism}}\ (\bibinfo  {publisher}
  {Wiley},\ \bibinfo {year} {1993})\ Chap.\ \bibinfo {chapter} {Magnetic
  Theory}, p.\ \bibinfo {pages} {431}\BibitemShut {NoStop}%
\bibitem [{\citenamefont {Bernas}\ and\ \citenamefont
  {Gabriel}(1973)}]{Bernas1973}%
  \BibitemOpen
  \bibfield  {author} {\bibinfo {author} {\bibfnamefont {H.}~\bibnamefont
  {Bernas}}\ and\ \bibinfo {author} {\bibfnamefont {H.}~\bibnamefont
  {Gabriel}},\ }\bibfield  {title} {\bibinfo {title} {Experimental and
  theoretical study of perturbed angular correlations for a rare-earth
  localized moment in iron},\ }\href@noop {} {\bibfield  {journal} {\bibinfo
  {journal} {Phys. Rev. B}\ }\textbf {\bibinfo {volume} {7}},\ \bibinfo {pages}
  {468} (\bibinfo {year} {1973})}\BibitemShut {NoStop}%
\bibitem [{\citenamefont {Fahlander}\ \emph
  {et~al.}(1979{\natexlab{b}})\citenamefont {Fahlander}, \citenamefont
  {Johansson}, \citenamefont {Lindgren},\ and\ \citenamefont
  {Possnert}}]{Fahlander1979}%
  \BibitemOpen
  \bibfield  {author} {\bibinfo {author} {\bibfnamefont {C.}~\bibnamefont
  {Fahlander}}, \bibinfo {author} {\bibfnamefont {K.}~\bibnamefont
  {Johansson}}, \bibinfo {author} {\bibfnamefont {B.}~\bibnamefont
  {Lindgren}},\ and\ \bibinfo {author} {\bibfnamefont {G.}~\bibnamefont
  {Possnert}},\ }\bibfield  {title} {\bibinfo {title} {Temperature dependence
  of the hyperfine fields for fluorine in ferromagnetic iron, nickel and
  gadolinium},\ }\href@noop {} {\bibfield  {journal} {\bibinfo  {journal}
  {Hyperfine Interactions}\ }\textbf {\bibinfo {volume} {7}},\ \bibinfo {pages}
  {299} (\bibinfo {year} {1979}{\natexlab{b}})}\BibitemShut {NoStop}%
\bibitem [{\citenamefont {Christiansen}\ \emph {et~al.}(1970)\citenamefont
  {Christiansen}, \citenamefont {Mahnke}, \citenamefont {Recknagel},
  \citenamefont {Riegel}, \citenamefont {Schatz}, \citenamefont {Weyer},\ and\
  \citenamefont {Witthuhn}}]{Christiansen1970}%
  \BibitemOpen
  \bibfield  {author} {\bibinfo {author} {\bibfnamefont {J.}~\bibnamefont
  {Christiansen}}, \bibinfo {author} {\bibfnamefont {H.~E.}\ \bibnamefont
  {Mahnke}}, \bibinfo {author} {\bibfnamefont {E.}~\bibnamefont {Recknagel}},
  \bibinfo {author} {\bibfnamefont {D.}~\bibnamefont {Riegel}}, \bibinfo
  {author} {\bibfnamefont {G.}~\bibnamefont {Schatz}}, \bibinfo {author}
  {\bibfnamefont {G.}~\bibnamefont {Weyer}},\ and\ \bibinfo {author}
  {\bibfnamefont {W.}~\bibnamefont {Witthuhn}},\ }\bibfield  {title} {\bibinfo
  {title} {Stroboscopic observation of nuclear larmor precession},\ }\href@noop
  {} {\bibfield  {journal} {\bibinfo  {journal} {Phys. Rev. C}\ }\textbf
  {\bibinfo {volume} {1}},\ \bibinfo {pages} {613} (\bibinfo {year}
  {1970})}\BibitemShut {NoStop}%
\bibitem [{\citenamefont {Georgiev}\ \emph {et~al.}(2006)\citenamefont
  {Georgiev}, \citenamefont {Matea}, \citenamefont {Balabanski} \emph
  {et~al.}}]{Georgiev2006}%
  \BibitemOpen
  \bibfield  {author} {\bibinfo {author} {\bibfnamefont {G.}~\bibnamefont
  {Georgiev}}, \bibinfo {author} {\bibfnamefont {I.}~\bibnamefont {Matea}},
  \bibinfo {author} {\bibfnamefont {D.}~\bibnamefont {Balabanski}}, \emph
  {et~al.},\ }\bibfield  {title} {\bibinfo {title} {$g$-factor of the 9/2+
  isomeric state in 65ni from transfer reaction},\ }\href@noop {} {\bibfield
  {journal} {\bibinfo  {journal} {Eur. Phys. J. A}\ }\textbf {\bibinfo {volume}
  {30}},\ \bibinfo {pages} {351} (\bibinfo {year} {2006})}\BibitemShut
  {NoStop}%
\bibitem [{\citenamefont {Recknagel}(1974)}]{Recknagel1974}%
  \BibitemOpen
  \bibfield  {author} {\bibinfo {author} {\bibfnamefont {E.}~\bibnamefont
  {Recknagel}},\ }in\ \href@noop {} {\emph {\bibinfo {booktitle} {{Nuclear
  Spectroscopy and Reactions, Part C}}}},\ \bibinfo {editor} {edited by\
  \bibinfo {editor} {\bibfnamefont {J.}~\bibnamefont {{Cerny}}}}\ (\bibinfo
  {publisher} {{Academic Press}, {New York}},\ \bibinfo {year} {1974})\
  p.~\bibinfo {pages} {93}\BibitemShut {NoStop}%
\bibitem [{\citenamefont {Stuchbery}(2003)}]{Stuchbery2003}%
  \BibitemOpen
  \bibfield  {author} {\bibinfo {author} {\bibfnamefont {A.~E.}\ \bibnamefont
  {Stuchbery}},\ }\bibfield  {title} {\bibinfo {title} {$\gamma$-ray angular
  distributions and correlations after projectile-fragmentation reactions},\
  }\href@noop {} {\bibfield  {journal} {\bibinfo  {journal} {Nuclear Physics
  A}\ }\textbf {\bibinfo {volume} {723}},\ \bibinfo {pages} {69 } (\bibinfo
  {year} {2003})}\BibitemShut {NoStop}%
\bibitem [{\citenamefont {Yamazaki}(1967)}]{Yamazaki1967}%
  \BibitemOpen
  \bibfield  {author} {\bibinfo {author} {\bibfnamefont {T.}~\bibnamefont
  {Yamazaki}},\ }\bibfield  {title} {\bibinfo {title} {Tables of coefficients
  for angular distribution of gamma rays from aligned nuclei},\ }\href@noop {}
  {\bibfield  {journal} {\bibinfo  {journal} {Nuclear Data Sheets. Section A}\
  }\textbf {\bibinfo {volume} {3}},\ \bibinfo {pages} {1 } (\bibinfo {year}
  {1967})}\BibitemShut {NoStop}%
\bibitem [{\citenamefont {Zobel}\ \emph {et~al.}(1979)\citenamefont {Zobel},
  \citenamefont {Cleemann}, \citenamefont {Eberth}, \citenamefont {Neumann},\
  and\ \citenamefont {Wiehl}}]{Zobel1979}%
  \BibitemOpen
  \bibfield  {author} {\bibinfo {author} {\bibfnamefont {V.}~\bibnamefont
  {Zobel}}, \bibinfo {author} {\bibfnamefont {L.}~\bibnamefont {Cleemann}},
  \bibinfo {author} {\bibfnamefont {J.}~\bibnamefont {Eberth}}, \bibinfo
  {author} {\bibfnamefont {W.}~\bibnamefont {Neumann}},\ and\ \bibinfo {author}
  {\bibfnamefont {N.}~\bibnamefont {Wiehl}},\ }\bibfield  {title} {\bibinfo
  {title} {High-spin states of negative parity in $^{69}\mathrm{Ge}$},\
  }\href@noop {} {\bibfield  {journal} {\bibinfo  {journal} {Phys. Rev. C}\
  }\textbf {\bibinfo {volume} {19}},\ \bibinfo {pages} {811} (\bibinfo {year}
  {1979})}\BibitemShut {NoStop}%
\bibitem [{\citenamefont {Stuchbery}\ and\ \citenamefont
  {Robinson}(2002)}]{Stuchbery2002}%
  \BibitemOpen
  \bibfield  {author} {\bibinfo {author} {\bibfnamefont {A.~E.}\ \bibnamefont
  {Stuchbery}}\ and\ \bibinfo {author} {\bibfnamefont {M.~P.}\ \bibnamefont
  {Robinson}},\ }\bibfield  {title} {\bibinfo {title} {{Perturbed
  $\gamma$--$\gamma$ correlations from oriented nuclei and static moment
  measurements I : formalism and principles}},\ }\href@noop {} {\bibfield
  {journal} {\bibinfo  {journal} {Nuclear Instruments and Methods in Physics
  Research A}\ }\textbf {\bibinfo {volume} {485}},\ \bibinfo {pages} {753}
  (\bibinfo {year} {2002})}\BibitemShut {NoStop}%
\bibitem [{\citenamefont {Carpenter}\ \emph {et~al.}(1990)\citenamefont
  {Carpenter}, \citenamefont {Bingham}, \citenamefont {Courtney}, \citenamefont
  {Janzen}, \citenamefont {Larabee}, \citenamefont {Liu}, \citenamefont
  {Riedinger}, \citenamefont {Schmitz}, \citenamefont {Bengtsson},
  \citenamefont {Bengtsson}, \citenamefont {Nazarewicz}, \citenamefont {Zhang},
  \citenamefont {Johansson}, \citenamefont {Popescu}, \citenamefont
  {Waddington}, \citenamefont {Baktash}, \citenamefont {Halbert}, \citenamefont
  {Johnson}, \citenamefont {Lee}, \citenamefont {Schutz}, \citenamefont
  {Nyberg}, \citenamefont {Johnson}, \citenamefont {Wyss}, \citenamefont
  {Dubuc}, \citenamefont {Kajrys}, \citenamefont {Monaro}, \citenamefont
  {Pilotte}, \citenamefont {Honkanen}, \citenamefont {Sarantites},\ and\
  \citenamefont {Haenni}}]{Carpenter1990}%
  \BibitemOpen
  \bibfield  {author} {\bibinfo {author} {\bibfnamefont {M.~P.}\ \bibnamefont
  {Carpenter}}, \bibinfo {author} {\bibfnamefont {C.~R.}\ \bibnamefont
  {Bingham}}, \bibinfo {author} {\bibfnamefont {L.~H.}\ \bibnamefont
  {Courtney}}, \bibinfo {author} {\bibfnamefont {V.~P.}\ \bibnamefont
  {Janzen}}, \bibinfo {author} {\bibfnamefont {A.~J.}\ \bibnamefont {Larabee}},
  \bibinfo {author} {\bibfnamefont {Z.~M.}\ \bibnamefont {Liu}}, \bibinfo
  {author} {\bibfnamefont {L.~L.}\ \bibnamefont {Riedinger}}, \bibinfo {author}
  {\bibfnamefont {W.}~\bibnamefont {Schmitz}}, \bibinfo {author} {\bibfnamefont
  {R.}~\bibnamefont {Bengtsson}}, \bibinfo {author} {\bibfnamefont
  {T.}~\bibnamefont {Bengtsson}}, \bibinfo {author} {\bibfnamefont
  {W.}~\bibnamefont {Nazarewicz}}, \bibinfo {author} {\bibfnamefont {J.~Y.}\
  \bibnamefont {Zhang}}, \bibinfo {author} {\bibfnamefont {J.~K.}\ \bibnamefont
  {Johansson}}, \bibinfo {author} {\bibfnamefont {D.~G.}\ \bibnamefont
  {Popescu}}, \bibinfo {author} {\bibfnamefont {J.~C.}\ \bibnamefont
  {Waddington}}, \bibinfo {author} {\bibfnamefont {C.}~\bibnamefont {Baktash}},
  \bibinfo {author} {\bibfnamefont {M.~L.}\ \bibnamefont {Halbert}}, \bibinfo
  {author} {\bibfnamefont {N.~R.}\ \bibnamefont {Johnson}}, \bibinfo {author}
  {\bibfnamefont {I.~Y.}\ \bibnamefont {Lee}}, \bibinfo {author} {\bibfnamefont
  {Y.~S.}\ \bibnamefont {Schutz}}, \bibinfo {author} {\bibfnamefont
  {J.}~\bibnamefont {Nyberg}}, \bibinfo {author} {\bibfnamefont
  {A.}~\bibnamefont {Johnson}}, \bibinfo {author} {\bibfnamefont
  {R.}~\bibnamefont {Wyss}}, \bibinfo {author} {\bibfnamefont {J.}~\bibnamefont
  {Dubuc}}, \bibinfo {author} {\bibfnamefont {G.}~\bibnamefont {Kajrys}},
  \bibinfo {author} {\bibfnamefont {S.}~\bibnamefont {Monaro}}, \bibinfo
  {author} {\bibfnamefont {S.}~\bibnamefont {Pilotte}}, \bibinfo {author}
  {\bibfnamefont {K.}~\bibnamefont {Honkanen}}, \bibinfo {author}
  {\bibfnamefont {D.~G.}\ \bibnamefont {Sarantites}},\ and\ \bibinfo {author}
  {\bibfnamefont {D.~R.}\ \bibnamefont {Haenni}},\ }\bibfield  {title}
  {\bibinfo {title} {Alignment processes and shape variations in $^{184}$pt},\
  }\href@noop {} {\bibfield  {journal} {\bibinfo  {journal} {Nuclear Physics
  A}\ }\textbf {\bibinfo {volume} {513}},\ \bibinfo {pages} {125 } (\bibinfo
  {year} {1990})}\BibitemShut {NoStop}%
\bibitem [{\citenamefont {Grau}\ \emph {et~al.}(1974)\citenamefont {Grau},
  \citenamefont {Rickey}, \citenamefont {Smith}, \citenamefont {Simms},\ and\
  \citenamefont {Tesmer}}]{Grau1974}%
  \BibitemOpen
  \bibfield  {author} {\bibinfo {author} {\bibfnamefont {J.~A.}\ \bibnamefont
  {Grau}}, \bibinfo {author} {\bibfnamefont {F.~A.}\ \bibnamefont {Rickey}},
  \bibinfo {author} {\bibfnamefont {G.~J.}\ \bibnamefont {Smith}}, \bibinfo
  {author} {\bibfnamefont {P.~C.}\ \bibnamefont {Simms}},\ and\ \bibinfo
  {author} {\bibfnamefont {J.~R.}\ \bibnamefont {Tesmer}},\ }\bibfield  {title}
  {\bibinfo {title} {The collective excitation of $^{103}$pd following (heavy
  ion, xn) reactions},\ }\href@noop {} {\bibfield  {journal} {\bibinfo
  {journal} {Nuclear Physics A}\ }\textbf {\bibinfo {volume} {229}},\ \bibinfo
  {pages} {346 } (\bibinfo {year} {1974})}\BibitemShut {NoStop}%
\bibitem [{\citenamefont {Simms}\ \emph {et~al.}(1974)\citenamefont {Simms},
  \citenamefont {Smith}, \citenamefont {Rickey}, \citenamefont {Grau},
  \citenamefont {Tesmer},\ and\ \citenamefont {Steffen}}]{Simms1974}%
  \BibitemOpen
  \bibfield  {author} {\bibinfo {author} {\bibfnamefont {P.~C.}\ \bibnamefont
  {Simms}}, \bibinfo {author} {\bibfnamefont {G.~J.}\ \bibnamefont {Smith}},
  \bibinfo {author} {\bibfnamefont {F.~A.}\ \bibnamefont {Rickey}}, \bibinfo
  {author} {\bibfnamefont {J.~A.}\ \bibnamefont {Grau}}, \bibinfo {author}
  {\bibfnamefont {J.~R.}\ \bibnamefont {Tesmer}},\ and\ \bibinfo {author}
  {\bibfnamefont {R.~M.}\ \bibnamefont {Steffen}},\ }\bibfield  {title}
  {\bibinfo {title} {Collective excitation of $^{101}\mathrm{Pd}$ following
  (heavy ion, $xn$) reactions},\ }\href@noop {} {\bibfield  {journal} {\bibinfo
   {journal} {Phys. Rev. C}\ }\textbf {\bibinfo {volume} {9}},\ \bibinfo
  {pages} {684} (\bibinfo {year} {1974})}\BibitemShut {NoStop}%
\bibitem [{\citenamefont {Kontani}\ \emph {et~al.}(1965)\citenamefont
  {Kontani}, \citenamefont {Asayama},\ and\ \citenamefont
  {Itoh}}]{Kontani1965}%
  \BibitemOpen
  \bibfield  {author} {\bibinfo {author} {\bibfnamefont {M.}~\bibnamefont
  {Kontani}}, \bibinfo {author} {\bibfnamefont {K.}~\bibnamefont {Asayama}},\
  and\ \bibinfo {author} {\bibfnamefont {J.}~\bibnamefont {Itoh}},\ }\bibfield
  {title} {\bibinfo {title} {Internal fields at nuclei of several impurities in
  ferromagnetic fe, co and ni alloys},\ }\href@noop {} {\bibfield  {journal}
  {\bibinfo  {journal} {Journal of the Physical Society of Japan}\ }\textbf
  {\bibinfo {volume} {20}},\ \bibinfo {pages} {1737} (\bibinfo {year}
  {1965})}\BibitemShut {NoStop}%
\bibitem [{\citenamefont {Georgiev}\ \emph {et~al.}(2002)\citenamefont
  {Georgiev}, \citenamefont {Neyens}, \citenamefont {Hass}, \citenamefont
  {Balabanski}, \citenamefont {Bingham}, \citenamefont {Borcea}, \citenamefont
  {Coulier}, \citenamefont {Coussement}, \citenamefont {Daugas}, \citenamefont
  {France}, \citenamefont {de~Oliveira~Santos}, \citenamefont {rska},
  \citenamefont {Grawe}, \citenamefont {Grzywacz}, \citenamefont {Lewitowicz},
  \citenamefont {Mach}, \citenamefont {Matea}, \citenamefont {Page},
  \citenamefont {tzner}, \citenamefont {Penionzhkevich}, \citenamefont {k},
  \citenamefont {Regan}, \citenamefont {Rykaczewski}, \citenamefont {Sawicka},
  \citenamefont {Smirnova}, \citenamefont {Sobolev}, \citenamefont {Stanoiu},
  \citenamefont {Teughels},\ and\ \citenamefont {Vyvey}}]{Georgiev2002}%
  \BibitemOpen
  \bibfield  {author} {\bibinfo {author} {\bibfnamefont {G.}~\bibnamefont
  {Georgiev}}, \bibinfo {author} {\bibfnamefont {G.}~\bibnamefont {Neyens}},
  \bibinfo {author} {\bibfnamefont {M.}~\bibnamefont {Hass}}, \bibinfo {author}
  {\bibfnamefont {D.~L.}\ \bibnamefont {Balabanski}}, \bibinfo {author}
  {\bibfnamefont {C.}~\bibnamefont {Bingham}}, \bibinfo {author} {\bibfnamefont
  {C.}~\bibnamefont {Borcea}}, \bibinfo {author} {\bibfnamefont
  {N.}~\bibnamefont {Coulier}}, \bibinfo {author} {\bibfnamefont
  {R.}~\bibnamefont {Coussement}}, \bibinfo {author} {\bibfnamefont {J.~M.}\
  \bibnamefont {Daugas}}, \bibinfo {author} {\bibfnamefont {G.~D.}\
  \bibnamefont {France}}, \bibinfo {author} {\bibfnamefont {F.}~\bibnamefont
  {de~Oliveira~Santos}}, \bibinfo {author} {\bibfnamefont {M.~G.}\ \bibnamefont
  {rska}}, \bibinfo {author} {\bibfnamefont {H.}~\bibnamefont {Grawe}},
  \bibinfo {author} {\bibfnamefont {R.}~\bibnamefont {Grzywacz}}, \bibinfo
  {author} {\bibfnamefont {M.}~\bibnamefont {Lewitowicz}}, \bibinfo {author}
  {\bibfnamefont {H.}~\bibnamefont {Mach}}, \bibinfo {author} {\bibfnamefont
  {I.}~\bibnamefont {Matea}}, \bibinfo {author} {\bibfnamefont {R.~D.}\
  \bibnamefont {Page}}, \bibinfo {author} {\bibfnamefont {M.~P.}\ \bibnamefont
  {tzner}}, \bibinfo {author} {\bibfnamefont {Y.~E.}\ \bibnamefont
  {Penionzhkevich}}, \bibinfo {author} {\bibfnamefont {Z.~P.}\ \bibnamefont
  {k}}, \bibinfo {author} {\bibfnamefont {P.~H.}\ \bibnamefont {Regan}},
  \bibinfo {author} {\bibfnamefont {K.}~\bibnamefont {Rykaczewski}}, \bibinfo
  {author} {\bibfnamefont {M.}~\bibnamefont {Sawicka}}, \bibinfo {author}
  {\bibfnamefont {N.~A.}\ \bibnamefont {Smirnova}}, \bibinfo {author}
  {\bibfnamefont {Y.~G.}\ \bibnamefont {Sobolev}}, \bibinfo {author}
  {\bibfnamefont {M.}~\bibnamefont {Stanoiu}}, \bibinfo {author} {\bibfnamefont
  {S.}~\bibnamefont {Teughels}},\ and\ \bibinfo {author} {\bibfnamefont
  {K.}~\bibnamefont {Vyvey}},\ }\bibfield  {title} {\bibinfo {title} {gfactor
  measurements of ~s isomeric states in neutron-rich nuclei around68ni produced
  in projectile-fragmentation reactions},\ }\href@noop {} {\bibfield  {journal}
  {\bibinfo  {journal} {Journal of Physics G: Nuclear and Particle Physics}\
  }\textbf {\bibinfo {volume} {28}},\ \bibinfo {pages} {2993} (\bibinfo {year}
  {2002})}\BibitemShut {NoStop}%
\bibitem [{\citenamefont {Bertschat}\ \emph {et~al.}(1972)\citenamefont
  {Bertschat}, \citenamefont {Haas}, \citenamefont {Leitz}, \citenamefont
  {Leithauser}, \citenamefont {Maier}, \citenamefont {Mahnke}, \citenamefont
  {Recknagel}, \citenamefont {Semmler}, \citenamefont {Sielemann},
  \citenamefont {Spellmeyer},\ and\ \citenamefont {Wichert}}]{Bertschat1972}%
  \BibitemOpen
  \bibfield  {author} {\bibinfo {author} {\bibfnamefont {H.}~\bibnamefont
  {Bertschat}}, \bibinfo {author} {\bibfnamefont {H.}~\bibnamefont {Haas}},
  \bibinfo {author} {\bibfnamefont {W.}~\bibnamefont {Leitz}}, \bibinfo
  {author} {\bibfnamefont {V.}~\bibnamefont {Leithauser}}, \bibinfo {author}
  {\bibfnamefont {K.~H.}\ \bibnamefont {Maier}}, \bibinfo {author}
  {\bibfnamefont {H.~E.}\ \bibnamefont {Mahnke}}, \bibinfo {author}
  {\bibfnamefont {E.}~\bibnamefont {Recknagel}}, \bibinfo {author}
  {\bibfnamefont {W.}~\bibnamefont {Semmler}}, \bibinfo {author} {\bibfnamefont
  {R.}~\bibnamefont {Sielemann}}, \bibinfo {author} {\bibfnamefont
  {B.}~\bibnamefont {Spellmeyer}},\ and\ \bibinfo {author} {\bibfnamefont
  {T.}~\bibnamefont {Wichert}},\ }\href@noop {} {\bibfield  {journal} {\bibinfo
   {journal} {Proceedings of International Conference on nuclear moments and
  nuclear structure, Osaka}\ ,\ \bibinfo {pages} {217}} (\bibinfo {year}
  {1972})}\BibitemShut {NoStop}%
\bibitem [{\citenamefont {Kr\'olas}(1974)}]{Krolas1974}%
  \BibitemOpen
  \bibfield  {author} {\bibinfo {author} {\bibfnamefont {K.}~\bibnamefont
  {Kr\'olas}},\ }\bibfield  {title} {\bibinfo {title} {Magnetic hyperfine
  fields at gallium in iron, cobalt, and nickel},\ }\href@noop {} {\bibfield
  {journal} {\bibinfo  {journal} {Proceedings of International Conference on
  Hyperfine Interactions}\ }\textbf {\bibinfo {volume} {4}},\ \bibinfo {pages}
  {151} (\bibinfo {year} {1974})}\BibitemShut {NoStop}%
\bibitem [{\citenamefont {Mohanta}\ \emph {et~al.}(2013)\citenamefont
  {Mohanta}, \citenamefont {Davane},\ and\ \citenamefont
  {Mishra}}]{Mohanta2013}%
  \BibitemOpen
  \bibfield  {author} {\bibinfo {author} {\bibfnamefont {S.~K.}\ \bibnamefont
  {Mohanta}}, \bibinfo {author} {\bibfnamefont {S.~M.}\ \bibnamefont
  {Davane}},\ and\ \bibinfo {author} {\bibfnamefont {S.~N.}\ \bibnamefont
  {Mishra}},\ }\bibfield  {title} {\bibinfo {title} {Tdpad measurement of
  magnetic hyperfine field for 66ga and 71as in ferromagnetic gd and tb},\
  }\href@noop {} {\bibfield  {journal} {\bibinfo  {journal} {Hyperfine
  Interactions}\ }\textbf {\bibinfo {volume} {221}},\ \bibinfo {pages} {29}
  (\bibinfo {year} {2013})}\BibitemShut {NoStop}%
\bibitem [{\citenamefont {Sood}(1978)}]{Sood1978}%
  \BibitemOpen
  \bibfield  {author} {\bibinfo {author} {\bibfnamefont {D.}~\bibnamefont
  {Sood}},\ }\bibfield  {title} {\bibinfo {title} {Empirical rules for
  substitutionality in metastable surface alloys produced by ion
  implantation},\ }\href@noop {} {\bibfield  {journal} {\bibinfo  {journal}
  {Physics Letters A}\ }\textbf {\bibinfo {volume} {68}},\ \bibinfo {pages}
  {469 } (\bibinfo {year} {1978})}\BibitemShut {NoStop}%
\bibitem [{\citenamefont {Allred}(1961)}]{Allred1961}%
  \BibitemOpen
  \bibfield  {author} {\bibinfo {author} {\bibfnamefont {A.}~\bibnamefont
  {Allred}},\ }\bibfield  {title} {\bibinfo {title} {Electronegativity values
  from thermochemical data},\ }\href@noop {} {\bibfield  {journal} {\bibinfo
  {journal} {Journal of Inorganic and Nuclear Chemistry}\ }\textbf {\bibinfo
  {volume} {17}},\ \bibinfo {pages} {215 } (\bibinfo {year}
  {1961})}\BibitemShut {NoStop}%
\bibitem [{\citenamefont {Slater}(1964)}]{Slater1964}%
  \BibitemOpen
  \bibfield  {author} {\bibinfo {author} {\bibfnamefont {J.~C.}\ \bibnamefont
  {Slater}},\ }\bibfield  {title} {\bibinfo {title} {Atomic radii in
  crystals},\ }\href@noop {} {\bibfield  {journal} {\bibinfo  {journal} {The
  Journal of Chemical Physics}\ }\textbf {\bibinfo {volume} {41}},\ \bibinfo
  {pages} {3199} (\bibinfo {year} {1964})}\BibitemShut {NoStop}%
\bibitem [{\citenamefont {Golovko}\ \emph {et~al.}(2005)\citenamefont
  {Golovko}, \citenamefont {Kraev}, \citenamefont {Phalet}, \citenamefont
  {Severijns}, \citenamefont {Z\'akouck\'y}, \citenamefont {V\'enos},
  \citenamefont {Herzog}, \citenamefont {Tramm}, \citenamefont {Srnka},
  \citenamefont {Honusek}, \citenamefont {K\"oster}, \citenamefont {Delaur\'e},
  \citenamefont {Beck}, \citenamefont {Kozlov}, \citenamefont {Lindroth},\ and\
  \citenamefont {Coeck}}]{Golovko2005}%
  \BibitemOpen
  \bibfield  {author} {\bibinfo {author} {\bibfnamefont {V.~V.}\ \bibnamefont
  {Golovko}}, \bibinfo {author} {\bibfnamefont {I.~S.}\ \bibnamefont {Kraev}},
  \bibinfo {author} {\bibfnamefont {T.}~\bibnamefont {Phalet}}, \bibinfo
  {author} {\bibfnamefont {N.}~\bibnamefont {Severijns}}, \bibinfo {author}
  {\bibfnamefont {D.}~\bibnamefont {Z\'akouck\'y}}, \bibinfo {author}
  {\bibfnamefont {D.}~\bibnamefont {V\'enos}}, \bibinfo {author} {\bibfnamefont
  {P.}~\bibnamefont {Herzog}}, \bibinfo {author} {\bibfnamefont
  {C.}~\bibnamefont {Tramm}}, \bibinfo {author} {\bibfnamefont
  {D.}~\bibnamefont {Srnka}}, \bibinfo {author} {\bibfnamefont
  {M.}~\bibnamefont {Honusek}}, \bibinfo {author} {\bibfnamefont
  {U.}~\bibnamefont {K\"oster}}, \bibinfo {author} {\bibfnamefont
  {B.}~\bibnamefont {Delaur\'e}}, \bibinfo {author} {\bibfnamefont
  {M.}~\bibnamefont {Beck}}, \bibinfo {author} {\bibfnamefont {V.~Y.}\
  \bibnamefont {Kozlov}}, \bibinfo {author} {\bibfnamefont {A.}~\bibnamefont
  {Lindroth}},\ and\ \bibinfo {author} {\bibfnamefont {S.}~\bibnamefont
  {Coeck}},\ }\bibfield  {title} {\bibinfo {title} {{Nuclear magnetic moment of
  $^{69}\mathrm{As}$ from on-line $\beta$-NMR on oriented nuclei}},\
  }\href@noop {} {\bibfield  {journal} {\bibinfo  {journal} {Phys. Rev. C}\
  }\textbf {\bibinfo {volume} {72}},\ \bibinfo {pages} {064316} (\bibinfo
  {year} {2005})}\BibitemShut {NoStop}%
\end{thebibliography}%

\end{document}